\documentclass[iop]{emulateapj}

\slugcomment{Accepted for publication in ApJ.}

\shorttitle{The central kiloparsec region of NGC\,1614
}

\shortauthors{Herrero-Illana et al.}

\usepackage{natbib}
\usepackage{amsmath,amssymb}

\def\asec{\ifmmode ^{\prime\prime}\else$^{\prime\prime}$\fi\,}
\def\degs{\ifmmode ^{\circ}\else$^{\circ}$\fi}

\def\ccsnrate{\mbox{\,$\nu_{\rm CCSN}$}}

\def\lir{\hbox{L$_{\rm IR}$}}
\def\lfir{\hbox{L$_{\rm FIR}$}}
\def\msun{\hbox{M$_{\odot}$}}

\newcommand{\Lsun}{\thinspace\hbox{$\hbox{L}_{\odot}$}}
\def\msunyr{\mbox{\,${\rm M_{\odot}\, yr^{-1}}$}}

\def\ergs{\mbox{\,erg~s$^{-1}$}}
\def\ergshz{\mbox{~erg~s$^{-1}$~Hz$^{-1}$}}

\def\lsim{\!\!\!\phantom{\le}\smash{\buildrel{}\over
 {\lower2.5dd\hbox{$\buildrel{\lower2dd\hbox{$\displaystyle<$}}\over
                                 \sim$}}}\,\,}
\def\gsim{\!\!\!\phantom{\ge}\smash{\buildrel{}\over
{\lower2.5dd\hbox{$\buildrel{\lower2dd\hbox{$\displaystyle>$}}\over
                               \sim$}}}\,\,}
\newcommand{\EE}[1]{\hbox{$\times 10^{ #1 }$}}

\begin{document}



\title{A multi-wavelength view of the central kiloparsec region in the Luminous Infrared Galaxy NGC\,1614
}


\author{Rub\'en Herrero-Illana\altaffilmark{1},
Miguel \'A. P\'erez-Torres\altaffilmark{1,2,3}, 
Almudena Alonso-Herrero\altaffilmark{4,5}, Antxon Alberdi\altaffilmark{1}, Luis Colina\altaffilmark{6}, Andreas Efstathiou\altaffilmark{7}, Lorena Hern\'andez-Garc\'ia\altaffilmark{1}, Daniel Miralles-Caballero\altaffilmark{8}, Petri V\"ais\"anen\altaffilmark{9,10}, 
Christopher C. Packham\altaffilmark{11},
Vinesh Rajpaul\altaffilmark{12},
and Albert A. Zijlstra\altaffilmark{13}}
\affil{$^1$Instituto de Astrof\'isica de Andaluc\'ia - CSIC, PO Box 3004, 18008, Granada, Spain\\
$^2$Centro de Estudios de F\'isica del Cosmos de Arag\'on (CEFCA), E-44001, Teruel, Spain\\
$^3$Dpto. de F\'isica Te\'orica, Universidad de Zaragoza, E-50009, Zaragoza, Spain \\
$^4$ Instituto de F\'isica de Cantabria, CSIC-Universidad de Cantabria, 39005, Santander, Spain\\
$^6$ Centro de Astrobiolog\'ia (INTA-CSIC), Ctra. de Torrej\'on a Ajalvir, km 4, 28850, Torrej\'on de Ardoz, Madrid, Spain\\
$^7$ School of Sciencies, European University Cyprus, Diogenes Street, Engomi, 1516, Nicosia, Cyprus\\
$^8$ Instituto de F\'isica Te\'orica, Universidad Aut\'onoma de Madrid, 28049, Madrid, Spain\\
$^9$ South African Astronomical Observatory, P.O. Box 9, Observatory 7935, Cape Town, South Africa\\
$^{10}$ Southern African Large Telescope, P.O. Box 9, Observatory 7935, Cape Town, South Africa\\
$^{11}$ Department of Physics and Astronomy, University of Texas at San Antonio, One UTSA Circle, San Antonio, TX 78249, USA\\
$^{12}$ Department of Physics, University of Oxford, Denys Wilkinson Building, Keble Road, Oxford OX1 3RH, UK\\
$^{13}$ Jodrell Bank Centre for Astrophysics, University of Manchester, Manchester M13 9PL, UK}


\altaffiltext{5}{Augusto Gonz\'alez Linares Senior Research Fellow.}


\begin{abstract}

The Luminous Infrared Galaxy  \object[NGC1614]{NGC\,1614} 
hosts a prominent circumnuclear ring of star formation.
However, the nature of the dominant emitting mechanism in its central $\sim100$\,pc is still under debate.
We present sub-arcsecond angular resolution radio, mid-infrared, Pa$\alpha$, optical, and X-ray observations of \object[NGC1614]{NGC\,1614}, aimed at studying in detail both the circumnuclear ring and the nuclear region.
The 8.4 GHz continuum emission traced by the Very Large Array (VLA) and the Gemini/T-ReCS 8.7 micron emission, as well as the Pa$\alpha$ line emission, show remarkable morphological similarities within the star-forming ring, suggesting that the underlying emission mechanisms are tightly related. 
We used an \emph{HST}/NICMOS Pa$\alpha$ map of similar resolution to our radio maps to disentangle the thermal free-free and non-thermal synchrotron radio emission, from which we obtained the intrinsic synchrotron power-law for each individual region within the central kpc of NGC\,1614. The radio ring surrounds a relatively faint, steep-spectrum source at the very center of the galaxy, suggesting that the central source is not powered by an AGN, but rather by a compact ($r \lsim 90$\,pc) starburst.
\textsl{Chandra} X-ray data also show that the central kpc region is dominated by starburst activity, without requiring the existence of an AGN.
We also used publicly available infrared data to model-fit  the spectral energy distribution of both the starburst ring and a putative AGN in NGC\,1614. In summary, we conclude that there is no need to invoke an AGN to explain the observed bolometric properties of the galaxy.
\end{abstract}


\section{Introduction}

Luminous Infrared Galaxies (LIRGs; $10^{11} \leq L_\mathrm{IR} / L_\odot \leq 10^{12}$;  
$L_\mathrm{IR}=L[8\textrm{ --- }1000\;\mu\mathrm{m}]$) are known to closely 
follow the far-infrared (FIR) to radio correlation, and hence must host either a burst of star formation, or an active galactic nucleus (AGN) at their center, or both. Disentangling 
whether an AGN, or a starburst is the dominant heating mechanism is essential to understand the role of the IR-phase in galaxy evolution \citep[see, e.g.][]{alonso-herrero12,alonso-herrero13}.
The central kpc regions of LIRGs are heavily enshrouded in dust, which prevents their study at optical wavelengths. 
Fortunately, at infrared wavelengths the extinction is significantly lower than in the optical, and radio is essentially extinction-free, which permits the study of the innermost regions of LIRGs, if the required spatial resolution is available.
Also, sub-arcsecond imaging with \emph{Chandra} allows to image the inner central regions  of LIRGs, thanks to the penetrating power of X-rays. 
Therefore, high-angular (sub-arcsecond) resolution observations of local ($D \lesssim 100$~Mpc) LIRGs at infrared, radio, and X-rays can be efficiently used to disentangle a putative AGN from a starburst.

NGC\,1614 (IRAS\, 04315-0840) is  a galaxy merger in a late stage of interaction, with strong tidal tails and only one obvious nucleus, though there is evidence for the remnant of a secondary one \citep{neff90, vaisanen12}. At a distance of 64\,Mpc (\citealt{devaucouleurs91}; 1\arcsec corresponds to 310\,pc), 
NGC\,1614 has an infrared luminosity $L_\mathrm{IR} \simeq 4\times10^{11} L_\odot$ 
\citep{sanders03}. 
Using the photometric information from the IRAS Bright Galaxy Sample \citep{sanders03} and from the NRAO VLA Sky Survey \citep{condon98}, the derived $q$-factor \citep{helou85}
is 2.46, well within the FIR-to-radio correlation. 

The central kpc region of NGC\,1614 hosts a prominent circumnuclear ring of star formation of $\sim600$\,pc diameter, revealed in Pa$\alpha$ \citep{alonso-herrero01}. Recently, several authors have suggested that the ring is formed by an inner Lindblad resonance, where the gas is driven to it through the dust lanes \citep{olsson10,konig13}.
The existence of an AGN at the center of NGC\,1614 is still a matter of debate: \citet{risaliti00} classified the hard X-ray emitting source in the central region of the galaxy as an AGN. However, the low signal-to-noise detection of the power-law continuum makes its interpretation uncertain \citep{olsson10}.
Sub-millimeter array (SMA) observations of NGC\,1614 seem to indicate a nuclear, non-thermal component, but which cannot be ascribed solely to an AGN, or to a starburst (SB) \citep{wilson08}.  
\citet{yuan10} have classified NGC\,1614 as a starburst-AGN composite (albeit with a significantly larger contribution from the starburst), using their new optical classification scheme.
More recently,  \citet{vaisanen12} have used 3.3$\,\mu$m spatially resolved polycyclic aromatic feature (PAH) imaging and continuum diagnostics to argue that an obscured AGN can be ruled out, concluding that NGC\,1614 is a pure starburst.

In this paper, we present sub-arcsecond angular resolution 
radio (3.6 and 6\,cm),  mid-IR (8.7$\mu$m),  optical (0.4 and 0.8$\mu$m), and \emph{Chandra} X-ray images of the central kpc region of NGC\,1614 to study in detail the central kpc region of this LIRG. Our main aim is to shed light on the AGN/SB controversy existing in the literature, as well as to 
discuss  the striking morphological similarities between the radio and mid-IR images, which suggest a common origin for both emission mechanisms.

\section{Observations and data reduction}

In this section, we  describe our sub-arcsecond resolution radio (3.6\,cm), mid-IR (8.7\,$\mu$m), PAH (3.3\,$\mu$m), Pa$\alpha$ (1.9\,$\mu$m) and optical (0.4 and 0.8\,$\mu$m) observations of NGC\,1614, which show striking morphological similarities (see Fig.~\ref{fig:all}), as well as \emph{Chandra} X-ray data and multi-epoch, archival VLA data at 3.6 and 6\,cm, aimed at characterizing the spectral index of the nucleus and the star-forming (SF) ring, as well as to study its variability (see Figs.~\ref{fig:variability} and \ref{fig:thermalfraction}).  
In Table~\ref{table:images}, we show the log for the radio and infrared observations discussed in this paper. 

\subsection{Radio}
We observed NGC\,1614 on November 2004 and May 2006 using the Karl G. Jansky Very Large Array (VLA) in A configuration at 3.6\,cm (project code AC749), in full polarization mode. The total synthesized bandwidth was of 100\,MHz. We used 3C\,48 as the absolute flux density calibrator, and 0420-014 (R.A.(J2000.0)=04$^\mathrm{h}$23$^\mathrm{m}$15$^\mathrm{s}$.801; Dec.(J2000.0)=$-$01\degs20'33.07\asec), at an angular distance of $7.7\degs$ from NGC\,1614, for phase-calibration purposes.
We used the NRAO {\it AIPS} package for all data reduction steps, including amplitude and phase calibration, as well as imaging.
We imaged the source using a pixel size of 0.05\asec and applied a natural weighting scheme, which resulted in a final synthesized beam of  $0.42\asec\times 0.25\asec$ at a position angle of $-33.8\degs$ for the observations on 2004, and a very similar one 
for the 2006 observations ($0.41\asec\times 0.24\asec$; $-28.47$\degs).
We used our VLA image on 2004 as the reference one, to compare with the infrared images. For this, we increased the pixel size of our radio images up to 0.089\asec.

To study the variability of the circumnuclear region in NGC\,1614, we also used publicly available VLA continuum data of NGC\,1614 in A-configuration obtained in May 1986 (project code AN37) at 6.0\,cm, and in July 1999 (project code AL503) at 3.6 and 6.0\,cm. 
We reduced these data in an analogous manner to that previously described for the 3.6\,cm data. We also used the same absolute flux density and phase calibrators mentioned above, thus ensuring that the astrometry was close to the milliarcsecond level. We obtained our images in the same manner, using an automatic script to prevent any  systematic errors.  In all cases, we used a natural weighting scheme for the \emph{u-v} data in the imaging process, using each time the same pixel size and convolving synthesized beam. 
The off-source r.m.s.\ noise attained in the final images was in good agreement with the expected theoretical thermal noise.

\begin{deluxetable*}{rrrrrrr}
\tabletypesize{\scriptsize}
\tablecaption{Observations summary}
\tablewidth{0pt}
\tablehead{
\colhead{} & \colhead{} & \colhead{} & \colhead{FWHM}  & \colhead{rms}  & \colhead{Peak} & \colhead{Astrometric shift} \\
\colhead{Date} & \colhead{Instrument} &\colhead{Wavelength} & \colhead{(arcsec)} & \colhead{($\mu$Jy)} & \colhead{(mJy)} & \colhead{(arcsec)} } 
\startdata
May 1986 &  VLA & 6.0\,cm & $0.71\times0.43$ &$65$& $4.91$ & 0 \\
Jul 1999 &  VLA & 6.0\,cm &  $0.60\times0.42$ &$30$& $4.46$  & 0 \\
Jul 1999 &  VLA & 3.6\,cm &  $0.50\times0.36$ &$35$& $2.65$  & 0 \\
Nov 2004 &  VLA & 3.6\,cm & $0.42\times0.25$ & $16$ & $1.63$  & 0 \\
May 2006 &  VLA & 3.6\,cm & $0.41\times0.24$ & $19$ & $1.51$  & 0 \\
\\
Feb 1998 & \emph{HST}/NICMOS & 1.9\,$\mu$m &  $0.15$ & $0.5$ & $0.19$  & $0.59$ \\
Aug 2006 & \emph{HST}/ACS & 0.4\,$\mu$m &  $0.10$ & $$1.0 & $0.02$  & $1.24$ \\
Aug 2006 & \emph{HST}/ACS & 0.8\,$\mu$m &  $0.10$ & $$1.9 & $0.04$  & $1.24$ \\
Sep 2006 & Gemini/T-ReCS & 8.7\,$\mu$m &  $0.38$ & $30$ & $1.64$ & $1.17$  \\
Jan 2009 & UKIRT/UIST & 3.3\,$\mu$m &  $0.33$ & $28$ & $0.22$ & $64.0$ 
\enddata
\label{table:images}
\tablecomments{The quoted angular resolution corresponds to the  diffraction-limited
FWHM for the radio and optical images, and is seeing-limited for the IR ones. The rms and peak flux density are given in units of mJy/beam and mJy/pixel for the radio and optical/IR images, respectively.
}
\end{deluxetable*}

\begin{deluxetable*}{rrrrrrrrr}
\tabletypesize{\scriptsize}
\tablecaption{Regions of enhanced radio and IR emission in NGC\,1614 and their integrated fluxes}
\tablewidth{0pt}
\tablehead{
\colhead{} & \colhead{Center} & \colhead{Diameter}  & \colhead{Diameter}  & \colhead{Pa$\alpha$} & \colhead{3.3\,$\mu$m} & \colhead{8.7\,$\mu$m} & \colhead{3.6\,cm} \\
\colhead{Region} & \colhead{(J2000)} & \colhead{(arcsec)} & \colhead{(pc)} & \colhead{(mJy)} & \colhead{(mJy)} & \colhead{(mJy)} & \colhead{(mJy)}} 
\startdata
A &  00.010s, 44.55\asec  &  $0.40$  & $124$ & $1.9\pm0.2 $   &   $1.5\pm0.2 $ & $21\pm2 $    & $ 1.41\pm0.05  $   \\
B &  59.990s, 45.11\asec  &  $0.86\times0.41$  & $266\times127$ & $3.9\pm0.4 $   &   $3.5\pm0.4 $ & $44 \pm4 $  & $2.95\pm0.12$   \\
C &  00.030s, 45.70\asec  &  $0.40$  & $124$ & $2.2\pm0.2 $   &   $1.5\pm0.2 $ & $19 \pm2 $   & $ 0.96\pm0.04  $   \\
D &  00.070s, 45.12\asec  &  $0.40$  & $124$ & $2.5\pm0.3 $   &   $1.4\pm0.2 $ & $22\pm2 $    & $ 1.07\pm0.04  $   \\
N &  00.027s, 45.12\asec  &  $0.60$  & $186$ & $4.3\pm0.4 $   &   $3.6\pm0.4 $ & $49 \pm5 $   & $ 1.70\pm0.08  $   \\
T &  00.027s, 45.12\asec  &  $2.50$  & $773$ & $47.1\pm4.7 $  &  $45.9\pm4.6 $ & $799\pm80 $   & $26.49\pm2.20 $ \\
R &  00.027s, 45.12\asec  &  T$-$N   & T$-$N & $42.8\pm4.3 $  &  $42.3\pm4.3 $ & $750 \pm75 $  & $24.78\pm2.10 $            
\enddata
\label{table:regionsandfluxes}
\tablecomments{Coordinates are given with respect to 4h34m, $-8$\degs34'. Region B is an ellipse with a position angle of 0\degs, for which we give major and minor axes. Region R is the ring defined by subtracting region N from total region T. The Pa$\alpha$ values are continuum-subtracted. The radio fluxes correspond to the image from November 2004, used as the reference radio image along the paper.}
\end{deluxetable*}

\subsection{Infrared}
We made use of the Gemini/T-ReCS $8.7\,\mu$m imaging data presented by \citet{diaz-santos08}, observed on 16-30 September 2006, using the T-ReCS instrument on the Gemini South telescope with a plate scale of $0.089\asec\,\mathrm{px}^{-1}$, with a total on-source integration time  of 1680 sec.
We refer the reader to \citet{diaz-santos08} for a detailed description of the observations and data reduction.

We also used Pa$\alpha$ observation obtained in February 1998 with the NICMOS camera on-board the \textit{Hubble Space Telescope} (\emph{HST}), using the filter F190N on NIC2, with a plate scale of $0.076\asec\,\mathrm{px}^{-1}$, and centered at $1.89\,\mu$m, for a total on-target time of 640 sec, presented in \citet{alonso-herrero01}. A nearby filter (F187N) was used to remove the stellar continuum.
We refer the reader to \citet{alonso-herrero00, alonso-herrero01} for details on the data reduction.

Finally, we made use of UIST Integral Field Unit observations on UKIRT performed in 6 and 26 January 2009, at 3.3\,$\mu$m with a total on-target exposure time of 4800\,s. The plate scale of the raw data was 0.12\asec\, px$^{-1}$ by 0.24\asec\, px$^{-1}$, yet the spatial resolution was similar to the IR data above.  The observations covered the 2.9 to 3.6\,$\mu$m range, and a continuum subtracted map of the 3.3\,$\mu$m PAH feature was specifically used here. For further details see \citet{vaisanen12}.

\subsection{Optical}
We  used publicly available optical images of NGC\,1614 taken in 2006 with the ACS camera, onboard the \emph{HST}. In particular, we retrieved broad-band images with filters F435W (centered at 4328.2 \AA{} $\sim 0.4$\,$\mu$m, $\sim$ B-band) and F814W (centered at 8057.0 \AA{} $\sim 0.8$\,$\mu$m, $\sim$ I-band). The pixel size of these images corresponds to $0.049\asec\,\mathrm{px}^{-1}$. 

We reduced the images using the on-the-fly \emph{HST} pipeline, applying the highest quality  reference files that were available at the time of retrieval. The calibrated F435W image had only few cosmic rays remaining.
However, the F814W image was severely contaminated with cosmic rays, which were removed as explained in \citet{miralles-caballero11}.

\subsection{X-Rays}
We used publicly available archival data of NGC\,1614 from ACIS-S, onboard \emph{Chandra}, taken on 21 November 2012. The pixel scale is $0.49\asec\,\mathrm{px}^{-1}$, and the spectral range covers from 0.2 to 10\,keV.
We used the CXC Chandra Interactive Analysis of Observations software package (CIAO\footnote{http://cxc.harvard.edu/ciao4.3/}), version 4.3, for the data reduction, which was made in the same way as described in \citet{hernandez-garcia13}.

We extracted a nuclear spectrum of NGC\,1614 from a 3$\arcsec$ aperture circular region centered at RA=4$^\mathrm{h}$34$^\mathrm{m}$0$^\mathrm{s}$.03, DEC=$-8$\degs34'45.\asec (J2000.0). We extracted the background spectrum from a 5$\arcsec$ aperture source-free circular region, in the same chip as the target, and close to the source, to minimize effects related to the spatial variations of the CCD response.

\subsection{Image alignment and estimation of flux density uncertainties}

We aligned the images by taking as reference position the flux density peak of the innermost region of our reference radio image (3.6\,cm map from November 2004), and then shifted the peaks of our $8.7\,\mu$m and Pa$\alpha$ images so as to make them coincide with the radio peak, with an estimated uncertainty of $0.04\asec$.The $3.3\,\mu$m PAH map was aligned by matching the corresponding strong $3$--$4\,\mu$m continuum nucleus to the nucleus seen in the $8.7\,\mu$m and Pa-alpha maps, with an estimated uncertainty of $0.05\asec$.

\begin{figure*}\centering
\includegraphics[width=0.89\textwidth]{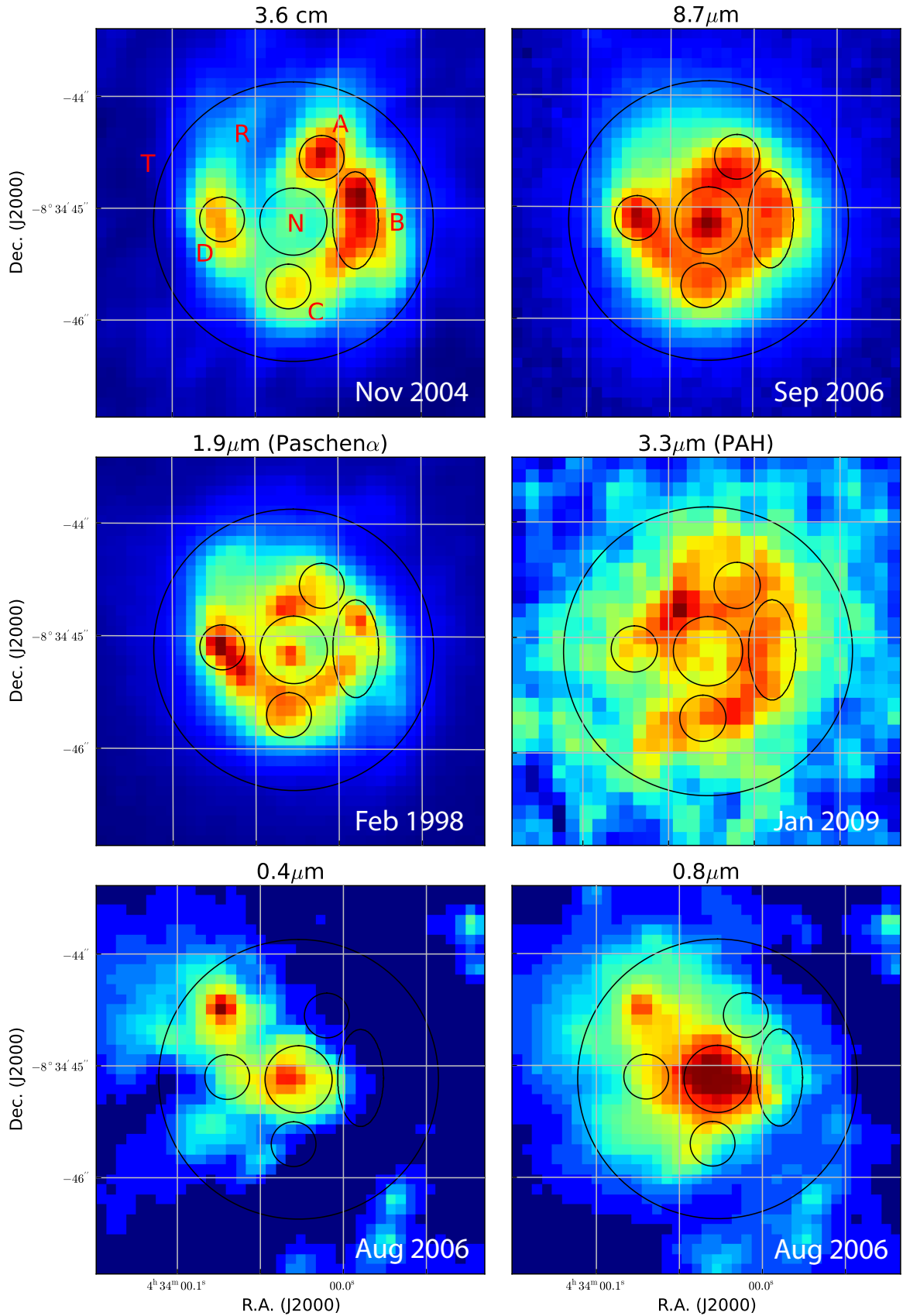}
\caption{Multi-wavelength comparison of the nuclear region of NGC\,1614. The top panels show radio (left) and mid-IR (right). 
The central panels show continuum-subtracted Paschen\,$\alpha$ (left) and the 3.3\,$\mu$m PAH feature (right). The bottom panels show \emph{HST}/ACS images at 0.4\,$\mu$m (left) and 0.8\,$\mu$m (right), B (F435W) and I (F814W) filters, respectively. Based on the radio emission, we have defined seven regions to compare the images, including the nuclear region (N), the whole ring (R) and the total area (T). Note the close similarity in the star formation ring and the contrast in the nuclear region between the radio and mid-IR wavelengths. Pixel sizes are the same for all images, except for the PAH feature map. Color scales are independent. See Fig.~2 in \citet{diaz-santos08} for a mid-IR/Pa$\alpha$ ratio image.}
\label{fig:all}
\end{figure*}

For the optical \emph{HST} images, which did not show a clear morphological correlation with the radio, or the infrared images, we used reference images at the intermediate wavelengths of 1.1\,$\mu$m and 1.6\,$\mu$m.
The resulting images are shown in Fig.~\ref{fig:all}. Table~\ref{table:images} shows the astrometric shift applied to the images with respect to the VLA astrometry.

To estimate the uncertainty in the (integrated) flux densities obtained for any given region, we 
used the following equation, which takes into account both thermal noise and systematic uncertainties, 

\begin{equation}
\sigma \simeq \sqrt{N_b\times\mathrm{rms}^2 + \left(\eta \times S_\mathrm{int}\right)^2},
\end{equation}
where rms is the off-source root mean square of the image; $\eta$ is a factor that accounts for uncertainties in the calibration system (we used $\eta=0.03$ for the VLA radio images and $\eta=0.10$ for the optical and IR images, which are conservative values); $S_\mathrm{int}$ is the integrated flux of the region of interest; and $N_b$ corresponds to  the number of beams that fits into a given region (the number of pixels in that region), for the radio (optical/IR) images. 

\section{Results and Discussion}

\subsection{Radio and infrared images}

We show in Figure~\ref{fig:all} our 3.6\,cm continuum VLA image of NGC\,1614 from November 2004 and the 8.7\,$\mu$m continuum T-ReCS image at similar angular resolution. We also show the \emph{HST}/NICMOS continuum-subtracted Pa$\alpha$ image
for comparison. 
The outer circle in the images covers essentially all of the emission at each wavelength, and has a radius of $\sim780$\,pc. 
We identify 
five regions within the circumnuclear ring:  A, B, C, and D, which correspond to areas of strong emission, and N, which  roughly delimits the nuclear region ($r \lesssim 90$\,pc).
For convenience, Figure~\ref{fig:all} shows two additional regions: R, which corresponds to the whole ring, and T, which encompasses the entire region (ring and nucleus, R+N). We show the locus and size for each of those regions, as well as their integrated 3.6\,cm, 8.7\,$\mu$m, 3.3\,$\mu$m and Pa$\alpha$ fluxes, in Table \ref{table:regionsandfluxes}.

The most conspicuous feature is the prominent mid-IR emission of the nucleus, N, which contrasts with its rather faint emission at radio wavelengths (see Fig.~\ref{fig:all} and Table~\ref{table:regionsandfluxes}). 
  The regions to the northwest of the ring, A and B, show a mid-IR/radio ratio below the average of the whole ring (R), while regions C and D show the opposite behavior.

\begin{figure*}\centering
\includegraphics[width=0.9\textwidth]{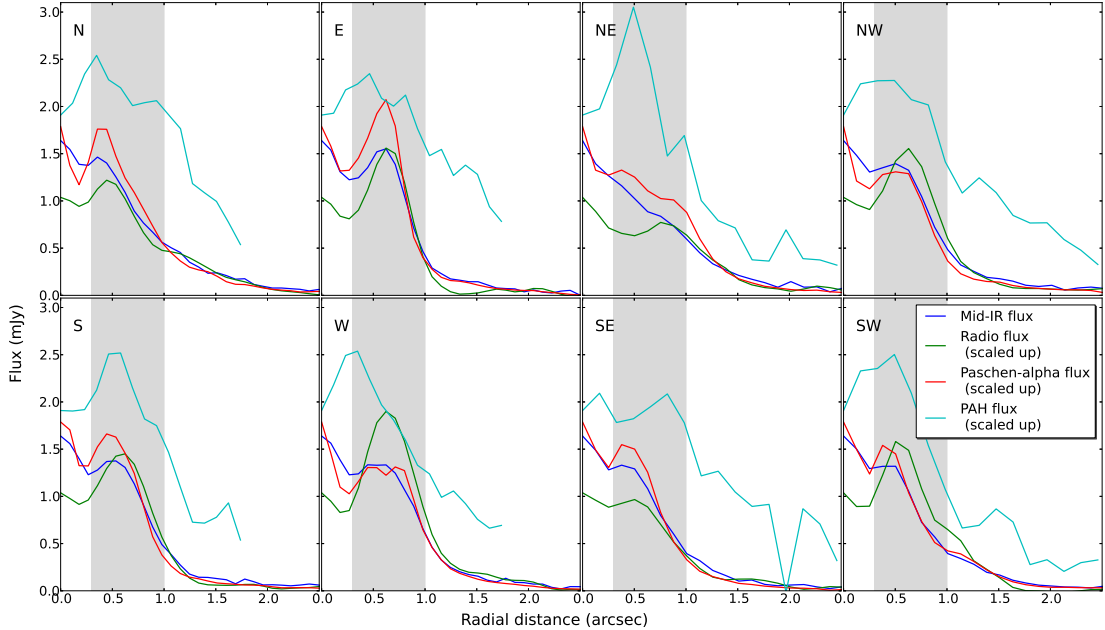}
\caption{Azimuthal profile of the fluxes in mid-IR (8.7\,$\mu$m, in blue), radio (3.6\,cm from Nov 2004, in green), Pa-$\alpha$ (1.9\,$\mu$m, in red) and continuum-subtracted PAH feature (3.3\,$\mu$m, in cyan) , starting from the center of the images. Radio, Pa-$\alpha$ and PAH fluxes are scaled up by the median of the 8.7\,$\mu$m/3.6\,cm, 8.7\,$\mu$m/Pa$\alpha$ and 8.7\,$\mu$m/3.3\,$\mu$m ratios, respectively (factors 19.2, 10.8 and 13.6). The profile is shown for eight cardinal directions, with an azimuthal binning of the size of a pixel (0.116\asec for the $3.3\,\mu$m image and 0.089\asec for the other cases). The shaded area, from 0.3 to 1.0\,arcsec, corresponds to the approximate width of the star formation ring.}
\label{fig:azimuthal}
\end{figure*}

Figure~\ref{fig:azimuthal} shows the azimuthal profiles at all three wavelengths, starting from the central pixel (the brightest pixel in region N for the 3.6\,cm and $8.4\,\mu$m images) and towards eight cardinal directions, separated by 45\degs\ from each other. 
To adequately compare the profiles, we normalized the radio continuum and Pa$\alpha$ fluxes to the
median of the ratios (8.7\,$\mu$m/3.6\,cm) of each region for the case of the mid-IR image, and to the median of the ratios (Pa$\alpha$/3.6\,cm) for the Pa$\alpha$ image.
We therefore increased the values of the 3.6\,cm and Pa$\alpha$ values by factors of $19.26$ and $10.79$, respectively, which allows to see more clearly variations in the whole  circumnuclear region.

Figure~\ref{fig:azimuthal} also shows that both the continuum 8.7\,$\mu$m  and Pa$\alpha$
emission follow almost exactly the same trend from the very center up to the outermost regions of the star-forming ring, as found by \citet{diaz-santos08}. Overall, the same trend is also seen when comparing the IR azimuthal profiles against those of the radio continuum. 
The remarkable morphological similarities seen at both radio and mid-infrared wavelengths in most regions of the circumnuclear ring strongly suggest that the mechanisms responsible for that emission must be related. \citet{konig13} show a similar plot (see their Fig.~6), with the azimuthal profile of the CO\,(2--1) and Pa$\alpha$ emission displayed together with the radio emission at three different bands.
The peaks of radio and mid-infrared emission in the ring would pinpoint then the regions where most of the starburst activity has taken place in the last 10-20 Myr.
The 8.7\,$\mu$m mid-IR flux includes both warm dust continuum and PAH emission.  The PAH emission is known to vary quite significantly from one SF region to another, or one galaxy to another, depending on the exact physical conditions \citep[see, e.g.,][]{calzetti07, diaz-santos10}.

There are morphological differences between the $3.3\mu$m PAH feature map and the rest of the IR and radio images. \citet{imanishi13} have suggested that either an age differentiation, or different dust extinction between the regions may explain those differences.

\citet{olsson10} ascribed the radio emission at 3.6 and 6.0\,cm from NGC\,1614 mainly to free-free emission from H\,{\sc ii} regions (mainly due to massive stars). As we shall show in the next sections, the non-thermal contribution from core-collapse supernovae (CCSNe) and supernova remnants (SNRs) is also significant within the circumnuclear ring, implying the existence of  young starbursts in the ring of NGC\,1614.

\begin{deluxetable*}{rrrrrrrrrr}
\tabletypesize{\scriptsize}
\tablecaption{Flux density and spectral index for the (circum)-nuclear region of NGC\,1614}
\tablewidth{0pt}
\tablehead{
\colhead{} &  \multicolumn{2}{c}{6.0\,cm} & \colhead{} & \multicolumn{3}{c}{3.6\,cm}   & \colhead{} & \multicolumn{2}{c}{Spectral index} \\
\cline{2-3}\cline{5-7}\cline{9-10} \\
\colhead{} & \colhead{1986}  & \colhead{1999} & \colhead{} & \colhead{1999} & \colhead{2004} & \colhead{2006}    & \colhead{} & \colhead{Total} & \colhead{Non-thermal}   \\   
\colhead{Region} & \colhead{(mJy)} & \colhead{(mJy)} & \colhead{} & \colhead{(mJy)} & \colhead{(mJy)}  & \colhead{(mJy)}  & \colhead{}  & \colhead{average} & \colhead{average} }
\startdata
A  & $  1.40 \pm 0.06 $  & $  1.46 \pm 0.05 $  & &  $  1.29 \pm 0.05 $  &$  1.26 \pm 0.04  $  &  $  1.23 \pm 0.04 $  &      & $-0.48\pm0.06 $ & $-0.56\pm0.10$  \\
B  & $  3.42 \pm 0.12 $  & $  3.71 \pm 0.11 $  & &  $  3.09 \pm 0.10 $  &$  3.01 \pm 0.09  $  &  $  2.88 \pm 0.09 $  &      & $-0.53\pm0.17 $ & $-0.72\pm0.16$  \\
C  & $  1.15 \pm 0.05 $  & $  1.23 \pm 0.04 $  & &  $  0.91 \pm 0.04 $  &$  1.00 \pm 0.03  $  &  $  0.97 \pm 0.03 $  &      & $-0.74\pm0.04 $ & $-1.44\pm0.13$  \\
D  & $  1.22 \pm 0.06 $  & $  1.13 \pm 0.04 $  & &  $  0.98 \pm 0.04 $  &$  1.08 \pm 0.04  $  &  $  1.00 \pm 0.04 $  &      & $-0.62\pm0.03 $ & $-1.14\pm0.22$  \\
N  & $  2.57 \pm 0.10 $  & $  2.81 \pm 0.09 $  & &  $  1.85 \pm 0.07 $  &$  1.79 \pm 0.06  $  &  $  1.79 \pm 0.06 $  &      & $-0.68\pm0.13 $ & $-1.30\pm0.53$  \\
T  & $ 34.06 \pm 1.06 $  & $ 35.70 \pm 1.08 $  & &  $ 25.61 \pm 0.79 $  &$ 26.39 \pm 0.80  $  &  $ 25.45 \pm 0.77 $  &      & $-0.73\pm0.25 $ & $-1.19\pm0.63$  \\
R  & $ 31.48 \pm 0.98 $  & $ 32.89 \pm 0.99 $  & &  $ 23.77 \pm 0.73 $  &$ 24.60 \pm 0.74  $  &  $ 23.66 \pm 0.72 $  &      & $-0.73\pm0.25 $ & $-1.18\pm0.63$  
\enddata
\label{table:variability}
\tablecomments{
Both images at 6.0\,cm were mapped with a beam of $0.71\times0.43$\,arcsec and the three epochs at 3.6\,cm with a beam of $0.50\times0.36$\,arcsec.
The  (two-point) spectral index was obtained for the 6.0 and 3.6\,cm simultaneous observations of NGC\,1614 carried out in 1999. Col. 7 shows the average spectral indices obtained directly from the spectral index map, while col. 8 spectral indices are obtained from the isolated non-thermal emission (see main text for details).}
\end{deluxetable*}

\subsection{Radio variability and spectral index of the circumnuclear ring in NGC\,1614}

\begin{figure*}
\includegraphics[width=\textwidth]{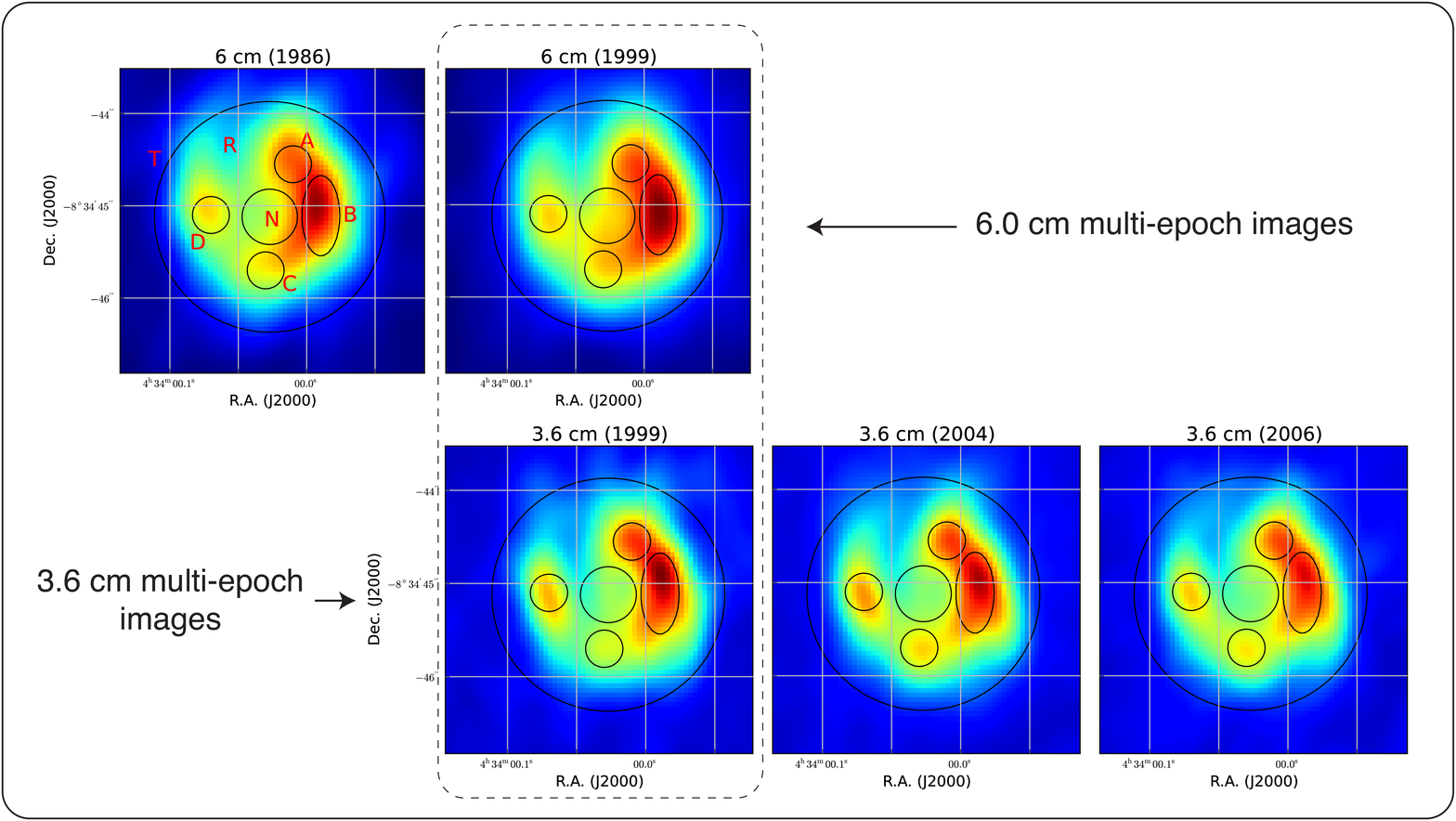}
\caption{Sub-arcsecond resolution radio continuum images of NGC\,1614. Top panels show images at a wavelength of 6\,cm from observations in 1986 (left) and 1999 (right). Bottom pannels show  3.6\,cm maps from observations in 1999 (left; quasi-simultaneously taken also at 6\,cm), 2004 (center) and 2006 (right). Images of the same frequency were mapped with the same beam. In all cases, the bulk of the radio emission is within a circumnuclear star-forming ring of radius $\sim390$\,pc. Note that the peaks of the brightest regions at 3.6\,cm have a maximum whose position coincides rather well with the peaks seen at 6\,cm, but for region B, whose maxima clearly peak at different positions. The color scale is the same for each epoch but independent for each wavelength, corresponding to [0.1, 5.0]\,mJy/beam for the 6\,cm band and [-0.17, 2.65]\,mJy/beam for 3.6\,cm. See main text and Table \ref{table:variability} for details.}
\label{fig:variability}
\end{figure*}

In star forming regions, non-thermal radio emission and variability are usually good tracers of recently exploded supernovae.
We show in Fig.~\ref{fig:variability} the radio interferometric images of NGC\,1614 at 6.0\,cm (epochs 1986 and 1999)
and 3.6\,cm (epochs 1999, 2004 and 2006), obtained with the VLA in A configuration, and imaged using the same restoring beam at all epochs, for consistency. In Table~\ref{table:variability}, we show the integrated flux for each defined region and for each epoch.

The images at each wavelength look overall similar, and the total radio emission at 3.6 and 6.0 cm has not varied among the epochs, 
within the uncertainties. In fact, while there seems to be an apparent increase of the flux density between 1986 and 1999 at 6\,cm (see Fig.~\ref{fig:variability}, top panel) for several regions, including N and B (the brightest spot in the circumnuclear ring), this variability is not quantitatively significant ($\lesssim1\sigma$, see Table~\ref{table:variability}) and hence  we cannot claim they are real. 
Similarly, the images at 3.6\,cm between 1996 and 2006 (Fig.~\ref{fig:variability}, bottom panel) suggest that the flux density in region B has been steadily decreasing from 1996 till 2006, while region  C would have experienced a rise and decrease of flux density during this period, possibly indicating supernova activity. Again, the variations are not significant, and unfortunately we lack further observations that could have allowed us to confirm, or rule out, those variations. 
We note, however, that the peaks of the regions sometimes show significant changes from epoch to epoch, suggesting that supernova events may be occurring.
Still, it seems that no very radio bright, Type IIn supernova ($L_{\nu, \mathrm{peak}} \gsim 10^{28}$\ergshz) has exploded in  NGC\,1614 during the 1986-2006 period. Such bright supernovae are known to evolve slowly and stay bright for $\gsim$10 yr, e.g. SN~1986J in NGC\,891 \citep{perez-torres02a}, or some of the supernovae in the compact nuclear starbursts of Arp~299-A \citep{perez-torres09b, bondi12} and Arp 220 \citep{parra07, batejat11}. Yet, we cannot exclude completely the possibility of a Type IIn SN having exploded and be missed by us, given the scarcity of the radio observations.

We used the quasi-simultaneous 3.6 and 6.0\,cm observations in 1999 to derive the average spectral index $\alpha$ ($S_\nu\propto\nu^\alpha$) for each of the regions  defined in Figs.~\ref{fig:all} and \ref{fig:variability}.  
The values in col.~7 of Table~\ref{table:variability} result from obtaining an average value of each region from the spectral index map created with the actual observations at 6.0 and 3.6\,cm from 1999 (which are the combination of the thermal free-free and non-thermal synchrotron radio emission), while col.~8 shows the intrinsic synchrotron radio spectral index, once the thermal component is subtracted from the total radio emission (see section 3.5 for details). 
The overall non-thermal spectral index for both the ring and the total area is $\alpha \simeq -1.2$. 
Such steep spectral index is suggestive of most of the diffuse, extended  synchrotron radio emission being due to supernovae and supernova remnants. Note, however, that the brightest regions in the ring have spectral indices significantly flatter than $-1.2$, which might suggest that the synchrotron radio emission is mostly powered by SN remnants, rather than by recently exploded SNe. 

The [Fe\,{\sc ii}] emission line at 1.26 and 1.64\,$\mu$m has also been proposed as a tracer of the supernova rate in nearby starburst galaxies \citep[see, e.g.,][]{moorwood88, greenhouse91, colina92, colina93, vanzi97, alonso-herrero03}, both in a pixel-by-pixel basis, and as an integrated approach.
The [Fe\,{\sc ii}] 1.26\,$\mu$m emission in NGC\,1614 shows a C-shape morphology in the circumnuclear star-forming ring \citep{rosenberg12}, with the brightest emitting region matching approximately region D in our 3.6 and 6.0\,cm images, i.e., an apparent anti-correlation between the maxima in the continuum VLA radio emission and the [Fe\,{\sc ii}] 1.26\,$\mu$m. However, after correcting for extinction the [Fe\,{\sc ii}] image, this anti-correlation disappears (Rosenberg, private communication).

\subsection{Spatially resolved X-ray emission}\label{sec:xray}
We fitted the \emph{Chandra} spectrum using standard procedures within the X-ray software
XSPEC\footnote{http://heasarc.gsfc.nasa.gov/xanadu/xspec/} version 12.7.0. We fitted the data using four different models:
(i) a pure thermal model (MEKAL), where the thermal emission is responsible for the bulk of the X-ray energy distribution;
(ii) an absorbed power-law model (PL), which corresponds to a non-thermal source representing an AGN; 
(iii) a composite of a thermal plus an absorbed power-law model (MEPL); and (iv) a thermal model with tuned individual abundances (VMEKAL), which models the metal abundance pattern of type II SNe \citep[see, e.g.,][]{iwasawa11, zaragoza-cardiel13}. In all models, we kept fixed the Galactic absorption to the predicted value using the {\sc nh} tool within {\sc ftools} \citep{dickley90, kalberla05}.

\begin{figure*}
\includegraphics[width=\textwidth]{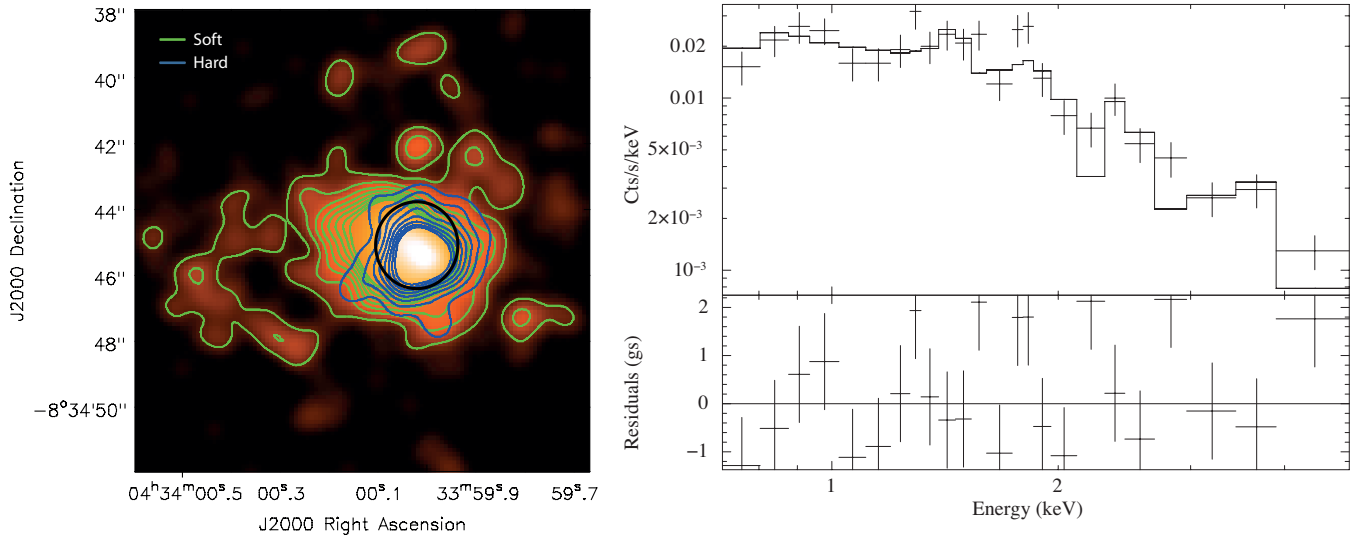}
\caption{\emph{Chandra} map (left) and spectrum fit (right) for NGC\,1614. The map shows the image of the total X-ray emission with the overlapped contours of the soft band (0.5--2.0\,keV, in green) and the hard band (2.0--10.0\,keV, in blue). The black circle corresponds to region T. Note that the emission is significantly more compact in the hard band than in the soft band.}
\label{fig:chandraall}
\end{figure*}

Neither the MEKAL nor the PL models yielded satisfactory fits to the data. The MEPL model fitted the data well, but with a physically unrealistic power-law index of $\Gamma=3.1$. Using a MEPL model to fit \emph{XMM-Newton} data (with a circular aperture of $\sim15\asec$), \citet{pereira-santaella11} obtained a luminosity four times higher, with $\Gamma=2$. 
Our best fit turned out to be the VMEKAL, which is shown in Fig.~\ref{fig:chandraall}, together with the soft and hard X-ray maps. This model finds abundances for Mg\,{\sc xi} (1.36\,keV), Si\,{\sc xiii} (1.85\,keV) and S\,{\sc xv} (2.4\,keV). The corresponding luminosity in the soft band for this model is $\log L\mathrm{(0.5-2 keV)}=40.78^{+0.04}_{-0.05}$\,erg/s, while there are not enough counts in the hard band to derive a realistic value for the luminosity. The fitted temperature is $kT=1.7\pm0.7$\,keV.

\subsection{Thermal free-free and non-thermal (synchrotron) radio emission}\label{sec:radioemission}

The bulk of the  continuum radio emission observed in the central region of NGC\,1614  comes from its circumnuclear ring (Fig.~\ref{fig:all}), where a strong burst of star-formation is ongoing \citep{alonso-herrero01}. Massive stars and their associated H\,{\sc ii} regions would be responsible for the thermal free-free radio emission, while supernovae and supernova remnants would account for the non-thermal synchrotron radio emission. Disentangling the contribution from each of those two components is non-trivial from radio measurements alone, since it would require observations at several frequencies in the range from  $\sim20$\,cm down to $\lsim$1\,cm, and ideally with the same angular resolution, which is not our case.

Here, we estimate the expected thermal free-free continuum radio emission from extinction corrected Pa$\alpha$ measurements. Then, using our (extinction-free) continuum radio data at 3.6\,cm from November 2004, we infer the amount of radio emission that is of non-thermal, synchrotron origin. 
As a bonus, from the Pa$\alpha$ measurements we obtain one of the most 
relevant physical parameters in the starburst in NGC\,1614, namely its Lyman photon ionizing flux, $N_\mathrm{ion}$, which is used in Section~\ref{sec:sedfit} to compare with the value derived from the SED fitting of the starburst in the circumnuclear region of NGC\,1614.
In fact, using standard relations \citep[see ][]{colina91} and the Pa$\alpha$ to H$\alpha$ recombination ratio and assuming no photon leakage,
we can derive the  ionizing photon flux, $N_\mathrm{ion}$, as:

\begin{equation}
N_\mathrm{ion} = 6.27\times10^{12}L_{\mathrm{Pa\alpha}} \ {\rm s^{-1}}, 
\end{equation}
where $N_\mathrm{ion}$ is measured in photons/s and $L_{\mathrm{Pa\alpha}}$ is the Pa$\alpha$ extinction-corrected luminosity,  in erg/s.
From our continuum-subtracted Pa$\alpha$ image,  we obtain a flux density of $\sim 47.1$\,mJy for the emission of the whole region, T, which corresponds to  an absorbed $L_{\rm Pa\alpha} = 3.6\times10^{41}\ergs$. This luminosity translates into an (absorbed) ionizing photon flux of $N_\mathrm{ion} \approx 2.27\EE{54} \ {\rm s^{-1}}$.  To obtain the relevant, unabsorbed ionizing photon flux, we corrected for the extinction,  $A_V$. Fortunately, the extinction towards NGC\,1614 is well studied, and is in the range $A_V = 3 - 5$ \citep[see, e.g.,][]{neff90, puxley94, alonso-herrero01, kotilainen01, rosenberg12}. Assuming a value of $A_V=4$, and using our observed Pa$\alpha$, we obtain the unabsorbed Pa$\alpha$ flux by using  standard H\,{\sc i} recombination lines ratios, for Case B \citep{baker38}, and using that

\begin{equation}
\frac{F(\lambda)}{I(\lambda)} = 10^{-C(H\beta)\left[f(\lambda)+1\right]},
\end{equation}
where $F(\lambda)$ and $I(\lambda)$ are the absorbed and unabsorbed fluxes, respectively, $C(H\beta)$ is the reddening coefficient, and $f(\lambda)$ is the reddening function \citep{cardelli89}.
The resulting unabsorbed Pa$\alpha$ flux is $\simeq80.2$ mJy, corresponding to an unabsorbed ionizing photon flux of $N_\mathrm{ion} \approx 3.87\EE{54} \ {\rm s^{-1}}$.

\begin{figure*}
\includegraphics[width=\textwidth]{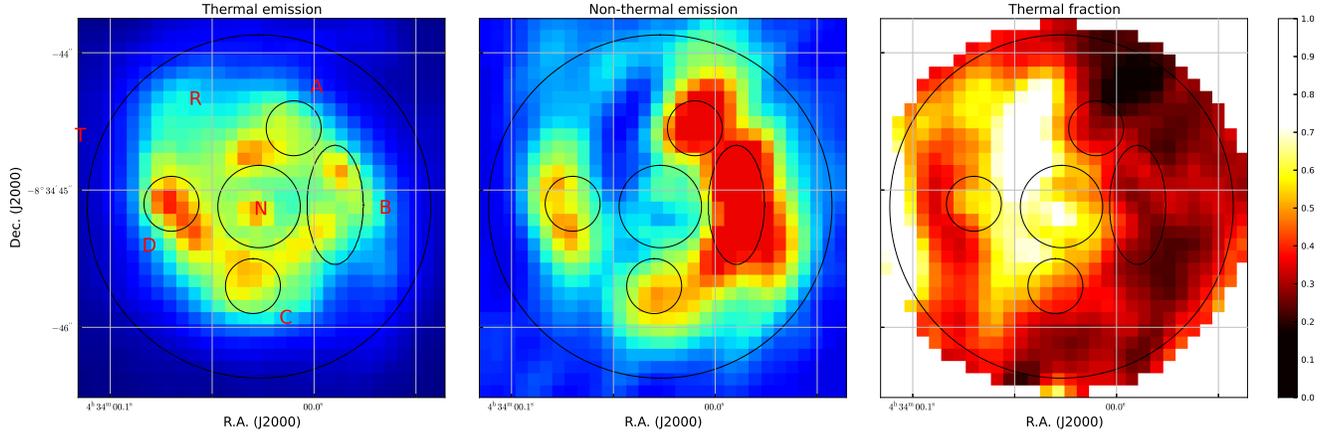}
\caption{Decomposition of the 3.6\,cm flux from November 2004 of the central region of NGC\,1614 into thermal (left panel, scaled version of the Pa$\alpha$ applying an uniform extinction of $A_V=4$) and non-thermal (middle panel, thermal radio emission subtracted from the total radio flux) components. The right panel shows the relative contribution of the thermal emission. Note that regions A and B are dominated by synchrotron non-thermal emission, in contrast with regions C and D.}
\label{fig:thermalfraction}
\end{figure*}

Using standard relations between  Pa$\alpha$ and H$\beta$ \citep[e.g.,][]{osterbrock89} and Eq.~3 from \citet{condon92}, we can obtain the 
thermal continuum radio emission as:

\begin{equation}
S_\mathrm{thermal} = 1.076\EE{13} \times F(\mathrm{Pa}\alpha) \,\nu^{-0.1},
\end{equation}
with $S_\mathrm{thermal}$ in mJy, the unabsorbed Pa$\alpha$ flux, $F(\mathrm{Pa}\alpha)$, in erg\,cm$^{-2}$\,s$^{-1}$, and $\nu$ in GHz, and
where we have assumed for simplicity a temperature of $10\,000$\,K and $N_\mathrm{e} = 10^4$\,cm$^{-3}$, which are typical of compact (i.e., size $\lesssim 1$\,pc) starburst regions. (The uncertainty in our estimates is dominated by the plasma electron temperature, since $S_{\rm th} \propto T_e^{0.52} $. A value of $T_e = 20000$ K would result in a thermal continuum radio flux $\sim 23$\% higher.)
In this way, we  isolated the thermal and non-thermal contributions to the radio emission, which are shown in Table~\ref{table:radioflux} and in Fig.~\ref{fig:thermalfraction}.
The corresponding thermal free-free radio flux density is 11.03\,mJy at 3.6\,cm for the whole  region, T, and is about 42\% of the total 3.6\,cm radio emission  in the central regions of NGC\,1614.
Since our Pa\,$\alpha$ measurements are much less affected by extinction than optical measurements, our decomposed values have a much weaker dependence on the actual value of the extinction. 
Indeed, allowing for an extinction $A_V$ in the range $(3-5)$, results in thermal free-free radio flux densities  in the range $(9.65-12.60)$\,mJy.

In summary, we obtained the thermal radio emission from a scaled version of the Pa$\alpha$ image, and the non-thermal radio emission as the result of the subtraction of the thermal emission from the total radio emission at 3.6\,cm from November 2004.
The ratio of thermal free-free to synchrotron radio emission can be used as an indicator of the starburst age of each region in the circumnuclear ring of NGC\,1614.  
Regions where there is essentially no synchrotron radio emission imply that supernovae have not yet started to explode, indicating ages of at most $\sim4$\,Myr, while regions where the supernovae have already started to explode would be older. 
The models of P\'erez-Olea \& Colina (1995) provide a quantitative estimate of the thermal free-free emission from massive stars, and non-thermal radio continuum emission from supernovae and supernova remnants. 
The ratios of thermal to non-thermal radio emission (see Table~\ref{table:radioflux} and Fig.~\ref{fig:thermalfraction}) for regions A and B are about $\sim0.5$, while those of regions C and D of 1.1 and 1.2, respectively. The ratios above (and the free-free thermal continuum luminosities) can be well explained if the emission in regions C and D come from instantaneous bursts with ages $\lsim$5.5 Myr, where supernovae have only recently started to explode. On the other hand, the emission from regions A and B would come from  slightly older ($\sim8$\,Myr) bursts, where essentially all exploding supernovae come from stars with masses in the 20-30 \msun\ range.

We note that the above discussion is valid at 3.6\,cm and is made under the assumption of a constant extinction of $A_V=4$ and variations across the ring may affect the thermal to non-thermal ratios. In fact, there seems to exist a gradient of the extinction increasing towards the west, as shown in Fig.~1 of \citet{konig13}. Considering an extinction of $A_V=5$ for regions A and B, and $A_V=3$ for regions C and D, we still obtain thermal to non-thermal radio emission ratios of $\sim0.6$ for regions A and B, $\sim0.7$ for region C, and $\sim0.8$ for region D.

The images of the decomposed radio emission help to understand the apparent paradox of the 
prominent mid-IR emission of the nucleus, N, which shows rather faint emission at radio wavelengths (see Fig.~\ref{fig:all} and Table~\ref{table:regionsandfluxes}). 
Since the thermal free-free emission is directly proportional to the Pa-$\alpha$ flux, the decomposed images would indicate that the radio emission from the nuclear region is dominated by thermal free-free emission (for any $A_V$ in the range $3-5$),  which in turn  suggests it is powered by a starburst, rather than an AGN, as we show in Section~\ref{sec:agn}.

In the absence of low-frequency absorption, we would expect that the emission at lower frequencies (e.g., 21\,cm) should be  dominated by synchrotron emission, in contrast to the 3.6 and 6.0\,cm images, where the contribution of the thermal emission is relevant. In fact, when we scale the non-thermal emission at 3.6\,cm 
(Fig.~\ref{fig:thermalfraction}) to 21\,cm, using our derived non-thermal spectral indices
(see Table~\ref{table:variability}), one would expect to recover the emission from the MERLIN image at 21\,cm shown in Fig.~4 in \citet{olsson10}.
Although the extrapolated image correlates, in general terms, well with the Olsson et al. image, all regions of NGC\,1614, except regions A and N, show at 21\,cm a lower flux density than expected, with the radio emission from region D being especially suppressed.
This should not come as a surprise, as those starburst regions have many massive stars that create big H\,{\sc ii} regions around them. Those H\,{\sc ii} regions are very efficient low-frequency absorbers, as demonstrated by, e.g., the large emission measure (EM) values in the vicinities of SN2000ft in the circumnuclear starburst of NGC 7469 \citep{alberdi06, perez-torres09a}, or around supernova A0 in Arp 299A \citep{perez-torres09b}. Indeed, the low-frequency absorption implies that region D, the one showing the least flux density at 21\,cm, has an EM $\approx 1.2 \times 10^7$ cm$^{-6}$\,pc, a value very similar to that found for in the vicinities of supernovae SN 2000ft in NGC7469, or A0 in Arp 299A. 
Regions B and C have  moderate EM values ($\approx [3.3, 4.7] \times 10^6$ cm$^{-6}$\,pc).
Finally, regions A and N have negligible EM values and essentially their radio emission is not being efficiently suppressed at low frequencies. 
While a discussion of the specific reasons for the differences in the low-frequency absorption displayed by the circumnuclear regions of NGC 1614 is beyond the scope of this paper, we just note here that our results are in agreement with all regions being synchrotron dominated at 21\,cm, but whose emission is being significantly suppressed in some regions by low-frequency absorption that is most likely due to foreground absorbers, i.e., H\,{\sc ii} regions. The main exceptions are regions A and N, which seems to suffer very little low-frequency absorption, possibly due to a smaller density in those region, as indicated by the low EM values.

\begin{deluxetable*}{rrrrrrr}
\tabletypesize{\scriptsize}
\tablecaption{Thermal and non-thermal radio emission in NGC\,1614}
\tablewidth{0pt}
\tablehead{
\colhead{} &  \colhead{$S_\mathrm{th}$} & \colhead{$S_\mathrm{syn}$}& \colhead{$L_\mathrm{th}$}&  \colhead{$L_\mathrm{syn}$}& \colhead{Unabs. $L_{\mathrm{Pa}\alpha}$}& \colhead{$N_\mathrm{ion}$}\\
\colhead{Region} &  \colhead{(mJy)} & \colhead{(mJy)} & \colhead{($10^{27}$ erg s$^{-1}$ Hz$^{-1}$)} &  \colhead{($10^{27}$ erg s$^{-1}$ Hz$^{-1}$)} & \colhead{($10^{40}$ erg s$^{-1}$ Hz$^{-1}$)}&  \colhead{($10^{53}$ s$^{-1}$)} \\
\colhead{(1)} & \colhead{(2)}& \colhead{(3)} & \colhead{(4)} & \colhead{(5)} & \colhead{(6)}& \colhead{(7)}  }
\startdata
A &  0.45   &  0.96  &   2.22 &    4.69 &  2.55  &   1.60   \\
B &  0.92   &  2.03  &   4.49 &    9.97 &  5.15  &   3.23   \\
C &  0.51   &  0.46  &   2.49 &    2.24 &  2.86  &   1.79   \\
D &  0.58   &  0.48  &   2.86 &    2.37 &  3.29  &   2.06   \\
N &  1.01   &  0.69  &   4.97 &    3.38 &  5.70  &   3.58   \\
T & 11.03   & 15.47  &  54.03 &   75.79 & 62.07  &  38.92   \\
R & 10.01   & 14.77  &  49.07 &   72.41 & 56.37  &  35.34 
\enddata
\label{table:radioflux}
\tablecomments{Col. 1: Region name; Col. 2: Thermal fraction of the flux at 3.6\,cm, obtained as a scaled version of the Pa$\alpha$ image assuming a constant extinction of $A_V=4$); Col. 3: Synchrotron fraction of the flux at 3.6\,cm, (obtained as Col.\,1 subtracted from the total radio emission); Col.4: Thermal radio luminosity at 3.6\,cm; Col. 5: Synchrotron radio luminosity at 3.6\,cm; Col. 6: Unabsorbed Pa$\alpha$ luminosity; Col. 7: Number of ionizing photons.}
\end{deluxetable*}

\begin{figure}\centering
\includegraphics[width=\columnwidth]{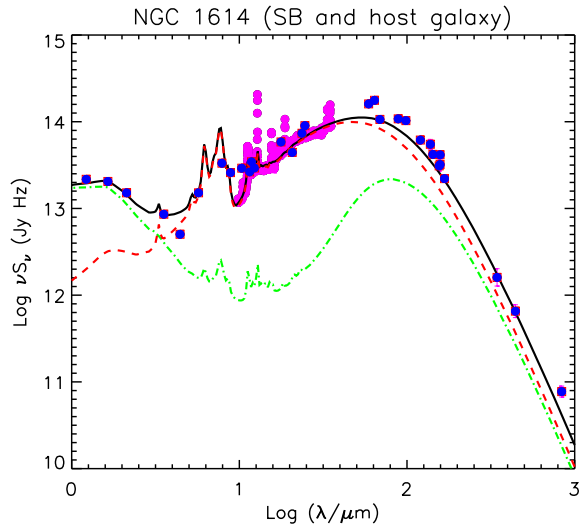}
\caption{NGC\,1614 SED fitting. Photometric data points are plotted in blue, while \emph{Spitzer} IRS spectrum is shown in pink. Lines show the overall fit (solid black), the starburst contribution (dashed red) and the host galaxy contribution (dot-dashed green).}
\label{fig:andreas}
\end{figure}

\begin{deluxetable}{rr}
\tabletypesize{\scriptsize}
\tablecaption{SED Model fitting}
\tablewidth{0pt}
\tablehead{
\colhead{Parameter} & \colhead{Value}} 
\startdata
$L_\mathrm{SB}$  				& $10^{11.39}L_\odot$   \\
e-folding time of SB 				& $35.4$\,Myr   \\
Age of SB 						& $29.5$\,Myr    \\
$\mathrm{SFR_\mathrm{max}}$ 		& $85.1 M_\odot\,\mathrm{yr}^{-1}$    \\
$\mathrm{SFR_\mathrm{mean}}$ 		& $57.7 M_\odot\,\mathrm{yr}^{-1}$   \\
								&	(averaged over 29.5\,Myr) \\
$\nu_\mathrm{SN}$ 				& $0.43\,\mathrm{yr}^{-1}$ \\
$N_\mathrm{ion}$					& $10^{54.54}\,\mathrm{s}^{-1}$
\enddata
\label{table:andreas}
\end{deluxetable}

\subsection{The star-formation and core-collapse supernova rates in NGC\,1614}\label{sec:sedfit}

The main goal of this section is to determine two of the most important parameters of any starburst, which are its star-formation rate (SFR) and its core-collapse supernova rate.  The (constant) CCSN rate, \ccsnrate, can be related to the (constant) $\mathrm{SFR}$ as follows \citep{perez-torres09a}:

\begin{equation}
\ccsnrate =  \mathrm{SFR}\, \left( \frac{\alpha-2}{\alpha-1} \right) \left(
          \frac{m_{\rm SN}^{1-\alpha} - m_u^{1-\alpha}}{m_l^{2-\alpha} 
           - m_u^{2-\alpha}} \right)
\label{eq:ccsnrate}
\end{equation}

where $\mathrm{SFR}$ is the (constant) star formation rate in \msunyr, $m_l$ and $m_u$ are the lower and upper mass limits of the initial mass function (IMF, $\Phi \propto m^{-\alpha}$), and $m_{\rm SN}$ is the minimum mass of stars that yield supernovae, assumed to be 8 \msun \citep[e.g.,][]{smartt09}. 
\citet{mattila01} found  an  empirical relationship between \lfir and \ccsnrate: \ccsnrate \ $\approx 2.7 \times 10^{-12}$\,(\lir/\Lsun)\,yr$^{-1}$. This implies a CCSN rate for the circumnuclear starburst of NGC\,1614 of $\approx1.08$\,SN\,yr$^{-1}$, for L$_{\rm IR} \simeq 4.0\times 10^{11} L_\odot$, which according to Eq. \ref{eq:ccsnrate} corresponds to a (constant) SFR of $\approx52.9\msunyr$.
However, a constant star-formation process  is likely to be, for LIRGs in general, and for NGC\,1614 in particular,  a poor approximation to the actual starburst scenario \citep{alonso-herrero01}. 

The opposite case to a constant SFR is that of a single instantaneous starburst.
For example, \citet{rosenberg12} used Starburst~99 \citep{leitherer99} to model the emission of NGC\,1614 within an instantaneous starburst scenario, and obtained an average age for the starburst of 6.4\,Myr and an integrated SN rate of 0.9\,yr$^{-1}$.
\citet{u12} found a star formation rate of $\mathrm{SFR}_\mathrm{UV+IR} \simeq 51.3\,M_\odot\,\mathrm{yr}^{-1}$. 
\citet{alonso-herrero01} modeled the star formation of NGC\,1614 using two Gaussian bursts, each of them with a FWHM of 5\,Myr, separated by 5\,Myr, obtaining an age of $\sim11$\,Myr after the peak of the first burst, i.e., a total age of $\sim16$\,Myr.  From the extinction corrected [Fe\,{\sc ii}], they predicted a supernova rate of 0.3\,SN/yr$^{-1}$.

An intermediate approach is that of an exponentially decaying starburst, which is the approach we have followed here.
Namely, we modeled the near-IR to sub-millimeter spectral energy distribution (SED) of NGC\,1614
by combining pure starburst models from \citet{efstathiou00}, revised by \citet{efstathiou09}, and models for the host galaxy. The latter models the emission from the stars using the models by \citet{bruzual03} and the emission from diffuse (cirrus) dust using the model of \citet{efstathiou09}. The fit is shown in Fig.~\ref{fig:andreas}, where photometric data points were obtained from NED\footnote{The NASA/IPAC Extragalactic Database (NED) is operated by the Jet Propulsion Laboratory, California Institute of Technology, under contract with the National Aeronautics and Space Administration.}, \citet{soifer01}, \citet{skrutskie06}, and \citet{ishihara10}.

The best fit starburst model yields a bolometric luminosity of $10^{11.39}L_\odot$, an initial star formation rate of 85.1 \msunyr (57.7 \msunyr averaged over the
duration of the starburst), a core-collapse supernova rate of 0.43\,SN\,yr$^{-1}$, and an ionizing photon flux of $3.47\times10^{54}$\,s$^{-1}$.

\subsection{Is there an AGN in the center of NGC\,1614?}\label{sec:agn}

The nuclear region, N, shows a non-thermal spectral index of $\alpha \sim -1.80$. 
Such a steep spectral index seems to be at odds with an AGN origin for the radio emission of region N,  but can be reconciled with the scenario of a compact starburst, powered by supernovae and supernova remnants. 
While such a high value of $\alpha$ is not rare or extreme for starbursts \citep[see, e.g., NGC\,253 in][]{heesen11}, it implies heavy synchrotron losses and, given the large radiation field in the nuclear region  (see the prominent 8.7\,$\mu$m continuum and Pa$\alpha$ line emission), also large inverse Compton losses. 
 We note that the radio spectral index is the average value over region N,  of $\sim90$\,pc in radius, so in principle we cannot rule out completely the existence of a hidden AGN inside that region, as found in other LIRGs, e.g., in Arp~299-A \citep{perez-torres10}. However, even if there is an AGN, its radio luminosity would not contribute more than $\sim6\%$ and $\sim 7.6\%$ at 3.6 and 6.0\,cm, respectively (see Table~\ref{table:variability}).
 For comparison, the AGN in Arp~299-A, as found from VLBI observations has a 5.0\,GHz flux density at cm-wavelenghts of about 820\,$\mu$Jy/b \citep{perez-torres10}, which corresponds to a luminosity of 
 $\nu\,L_{\nu, \rm AGN} \sim 9\times 10^{36}$\ergs, and accounts for no more than about 11\% of the compact VLBI 5.0 GHz flux. This luminosity value is also less than 1\% of the total 5.0 GHz emission, as traced by eMERLIN within the region where all SNe/SNRs are exploding \citep[see Fig.~4 and sect.~3.3 in][]{bondi12}. 
  The corresponding 5.0 GHz luminosity of a putative AGN in NGC\,1614 is therefore no more than
$\nu\,L_{\nu, \rm AGN} \sim 7.1\times 10^{37}$\ergs\ and, if most of this luminosity comes in turn from a 
nuclear starburst, the AGN must be even fainter.

The ratio of the 8.7\,$\mu$m/Pa$\alpha$ emission also suggests that region N is powered by a burst of star formation. In fact, the ratio in the central $\sim$90 pc agrees well with the ratios obtained for nuclei of H\,{\sc ii} systems, and is significantly lower than obtained for the nuclei of Sy/Sy 2 systems (Table 3 in D\'iaz-Santos et al. 2008), which are known to host an AGN. 

Similarly, the \emph{Chandra} X-ray emission supports a starburst driven scenario for the central regions of NGC\,1614.
We calculated hardness ratios, which are model-independent, for different apertures.
We defined the hardness ratio as $\mathrm{HR}=(H-S)(H+S)$, being $H$ and $S$ the hard [2.0--10.0]\,keV and soft [0.5--2.0]\,keV bands, respectively. For an aperture  of 3\asec in radius, we get HR$=-0.40$, while  for an aperture of 0.3\asec (coincident with region N) we get
a significantly harder spectrum, HR=+0.69. The point at which HR becomes positive (i.e., the hard emission dominates) is $\lsim0.4\asec$. 
The weak hard X-ray emission could easily be due to the presence of X-ray binaries in a compact starburst in the central 0.3\asec (i.e., $\sim110$\,pc), in agreement with the sizes of the starbursts seen in, e.g., Arp~299-A \citep{perez-torres09a, bondi12} and Arp~220 \citep{parra07}. While we cannot rule out completely that an AGN  makes also some contribution, the presence of emission lines of Mg\,{\sc xi}, Si\,{\sc xiii} and S\,{\sc xv} in the spectrum (see Fig.~\ref{fig:chandraall})
suggests the existence of SNe and/or SNRs. Although the temperature is somewhat higher than expected for SNRs, with typical values of kT=0.5\,keV \citep[see, e.g.,][]{soria03}, they are in good agreement with values obtained for young supernovae \citep[e.g., SN2001gd in NGC\,5033 had kT=1.1 keV;][]{perez-torres05}.

\begin{figure}
\includegraphics[width=\columnwidth]{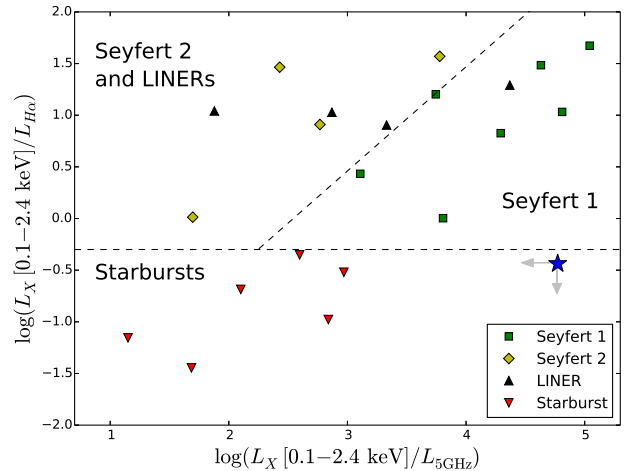}
\caption{Multi-wavelength diagnostic plot discriminating starbursts from AGN. NGC\,1614 is plotted as a blue star. Since the used X-ray aperture (3\asec) is larger than the H$\alpha$ and 5\,GHz ones (region N, 0.6\asec), NGC\,1614 real position in the diagram will necessarily move down and to the left, making NGC\,1614 fall clearly in the starburst dominated region. We derive an X-ray luminosity in the range [0.1--2.4]\,keV of $10^{40.90}$\,erg/s. Adapted from \citet{perez-olea96}}
\label{fig:diagnostic}
\end{figure}

Additionally, using the multi-wavelength optical, radio, and soft X-rays diagnostic diagram from \citet{perez-olea96}, region N would be in the starburst dominated region, as can be seen in Fig.~\ref{fig:diagnostic}, with $\log(L_X/L_\mathrm{5GHz}) < 4.77$ and $\log(L_X/L_\mathrm{H\alpha}) < -0.43$. (The inequality accounts for the fact that the radio and H$\alpha$ luminosity are for region N, while the X-ray luminosity corresponds to a 3\asec aperture.)

We also used archival data from \emph{Spitzer} IRS to check the high-resolution spectra for NGC\,1614 looking for [Ne\,{\sc v}] lines at $14.3$ and $24.3\,\mu$m, which would be indicative of the existence of an AGN \citep{genzel98, armus07}, but found no evidence of their presence.

Finally, we also fitted the multi-wavelength SED with a combination of starburst models and AGN torus models \citep{efstathiou95} and found that the contribution of the AGN to the total bolometric luminosity, if any, would be at most $\sim10\%$.

In summary, all evidence shows that the bulk of the observed emission (at all wavelengths) from the circumnuclear region of NGC\,1614 can be explained with the existence of a powerful starburst,  without any need to advocate the existence of an AGN.

\section{Summary}

We have presented sub-arsecond angular resolution radio, mid-infrared , optical, and X-ray observations of the central kiloparsec region of {NGC\,1614}. Our main results are as follows:

\begin{enumerate}
\item The continuum emission of the circumnuclear ring, as traced by 3.6\,cm Very Large Array (VLA), T-ReCS 8.7\,$\mu$m and \emph{HST}/NICMOS Pa$\alpha$ show remarkable morphological similarities, suggesting a common origin for both radio, mid-IR, and hydrogen recombination line emission in the ring, likely recent star-forming activity along the ring.  

\item The analysis of multi-epoch VLA observations at 3.6 and 6.0\,cm spanning almost 20 years show that the total radio luminosity of the ring is quite stable, and the possible variations are not significant. 
Still, if such variability was real, it could be ascribed to CCSN activity, since the luminosities involved in them ($L_{\nu} \sim (1.0-1.5)\times 10^{27}$\ergshz), are typical of Type IIb/IIL supernovae. 
The steady radio emission of NGC\,1614 is rather high in the whole circumnuclear ring, likely due to significant thermal free-free emission from massive stars and H\,{\sc ii} regions, as well as to diffusion of  synchrotron emission produced in the shocks of SNe and SNRs.

\item From the Pa$\alpha$ image of the circumnuclear ring, we measured the ionizing photon flux for the ring and the nucleus, and predicted the (thermal) free-free radio emission. We then estimated, from our radio images, the intrinsic non-thermal synchrotron contribution to the observed radio continuum.  We find that the circumnuclear (radio) ring surrounds a faint and compact ($r \lsim 90$\,pc) source at the very center of the galaxy, with a non-thermal steep spectrum ($\alpha_\mathrm{syn} \simeq -1.3$), suggesting that the central source is not powered by an AGN, but rather by a compact  starburst. In any case, a putative AGN would contribute at most $\sim8\%$ to the total radio luminosity.

\item We have modeled the X-ray spectrum of the central 3-kpc region of NGC\,1614, using \emph{Chandra} X-ray data. This region is well described by a pure thermal model with $k\,T \sim1.7$\,keV), with the presence of lines indicative of SNe/SNR. The circumnuclear region has a hardness ratio $\mathrm{HR}\simeq-0.40$, 
in good agreement with expectations for a starburst, while the inner $\sim 0.4$\asec\, ($\sim 120$\,pc) of NGC\,1614 shows HR$\simeq +0.69$, which could be explained by a population of high-mass X-ray  binaries. 

\item We also have used several diagnostic diagrams, which suggest that region N (the nuclear region) has no AGN inside \citep[e.g.,][]{perez-olea96, asmus11} and is dominated by a starburst.

\item Finally, we have used publicly available infrared data to perform a model-fit to the spectral energy distribution of NGC\,1614.
We find that the circumnuclear star-forming region in NGC\,1614 can be well described by an exponentially decaying burst that started $\lsim$30 Myr, and which has an average core-collapse supernova rate of $\sim0.4\,$ SN yr$^{-1}$ and an average SFR rate of $\sim58$\,\msunyr.   

\end{enumerate}

In summary, although a dust-enshrouded AGN cannot be completely ruled out by our observations, there is no need to advocate its existence, since the starburst completely dominates the observed properties of both the circumnuclear star-forming ring and the nuclear region.  We have proposed deep, dual-frequency, contemporaneous VLBI observations, which will unambiguously show whether there is a faint AGN sitting at the very center of NGC\,1614.

\acknowledgments

We thank the anonymous referee for his/her constructive comments which helped to improved our paper. We are grateful to Alba Fern\'andez-Mart\'in and Josefa Masegosa for useful discussions, and to Tanio D\'iaz-Santos for his help with the Pa$\alpha$ images of NGC\,1614.

RHI, MAPT, and AA acknowledge support by the Spanish MINECO through grants AYA 2009-13036-C02-01 and AYA 2012-38491- C02-02, cofunded with FEDER funds. AAH acknowledges support from the Universidad de Cantabria through the Augusto G. Linares program and the Spanish MINECO through grant AYA 2012-31447. LC acknowledges support by the Spanish MINECO through grant AYA 2010-21161-C02-01. PV acknowledges support from the National Research Foundation.

The National Radio Astronomy Observatory is a facility of the National Science Foundation operated under cooperative agreement by Associated Universities, Inc.

This research made use of data obtained from the \emph{Chandra} Data Archive provided by the \emph{Chandra} X-ray Center (CXC).

NGC\,1614 maps on this paper were produced using Python's \emph{Kapteyn Package} \citep{kapteyn12}.

{\it Facilities:} \facility{CXO}, \facility{Gemini:South (T-ReCS)}, \facility{HST (ACS, NICMOS)}, \facility{UKIRT:UIST}, \facility{VLA}.

\bibliographystyle{apj}
\bibliography{masterbib}

\begin{thebibliography}{67}
\expandafter\ifx\csname natexlab\endcsname\relax\def\natexlab#1{#1}\fi

\bibitem[{{Alberdi} {et~al.}(2006){Alberdi}, {Colina}, {Torrelles}, {Panagia},
  {Wilson}, \& {Garrington}}]{alberdi06}
{Alberdi}, A., {Colina}, L., {Torrelles}, J.~M., {et~al.} 2006, \apj, 638, 938

\bibitem[{{Alonso-Herrero} {et~al.}(2001){Alonso-Herrero}, {Engelbracht},
  {Rieke}, {Rieke}, \& {Quillen}}]{alonso-herrero01}
{Alonso-Herrero}, A., {Engelbracht}, C.~W., {Rieke}, M.~J., {Rieke}, G.~H., \&
  {Quillen}, A.~C. 2001, \apj, 546, 952

\bibitem[{{Alonso-Herrero} {et~al.}(2013){Alonso-Herrero}, {Pereira-Santaella},
  {Rieke}, {Diamond-Stanic}, {Wang}, {Hern{\'a}n-Caballero}, \&
  {Rigopoulou}}]{alonso-herrero13}
{Alonso-Herrero}, A., {Pereira-Santaella}, M., {Rieke}, G.~H., {et~al.} 2013,
  \apj, 765, 78

\bibitem[{{Alonso-Herrero} {et~al.}(2012){Alonso-Herrero}, {Pereira-Santaella},
  {Rieke}, \& {Rigopoulou}}]{alonso-herrero12}
{Alonso-Herrero}, A., {Pereira-Santaella}, M., {Rieke}, G.~H., \& {Rigopoulou},
  D. 2012, \apj, 744, 2

\bibitem[{{Alonso-Herrero} {et~al.}(2003){Alonso-Herrero}, {Rieke}, {Rieke}, \&
  {Kelly}}]{alonso-herrero03}
{Alonso-Herrero}, A., {Rieke}, G.~H., {Rieke}, M.~J., \& {Kelly}, D.~M. 2003,
  \aj, 125, 1210

\bibitem[{{Alonso-Herrero} {et~al.}(2000){Alonso-Herrero}, {Rieke}, {Rieke}, \&
  {Scoville}}]{alonso-herrero00}
{Alonso-Herrero}, A., {Rieke}, G.~H., {Rieke}, M.~J., \& {Scoville}, N.~Z.
  2000, \apj, 532, 845

\bibitem[{{Armus} {et~al.}(2007){Armus}, {Charmandaris}, {Bernard-Salas},
  {Spoon}, {Marshall}, {Higdon}, {Desai}, {Teplitz}, {Hao}, {Devost}, {Brandl},
  {Wu}, {Sloan}, {Soifer}, {Houck}, \& {Herter}}]{armus07}
{Armus}, L., {Charmandaris}, V., {Bernard-Salas}, J., {et~al.} 2007, \apj, 656,
  148

\bibitem[{{Asmus} {et~al.}(2011){Asmus}, {Gandhi}, {Smette}, {H{\"o}nig}, \&
  {Duschl}}]{asmus11}
{Asmus}, D., {Gandhi}, P., {Smette}, A., {H{\"o}nig}, S.~F., \& {Duschl}, W.~J.
  2011, \aap, 536, A36

\bibitem[{{Baker} \& {Menzel}(1938)}]{baker38}
{Baker}, J.~G., \& {Menzel}, D.~H. 1938, \apj, 88, 52

\bibitem[{{Batejat} {et~al.}(2011){Batejat}, {Conway}, {Hurley}, {Parra},
  {Diamond}, {Lonsdale}, \& {Lonsdale}}]{batejat11}
{Batejat}, F., {Conway}, J.~E., {Hurley}, R., {et~al.} 2011, \apj, 740, 95

\bibitem[{{Bondi} {et~al.}(2012){Bondi}, {P{\'e}rez-Torres}, {Herrero-Illana},
  \& {Alberdi}}]{bondi12}
{Bondi}, M., {P{\'e}rez-Torres}, M.~A., {Herrero-Illana}, R., \& {Alberdi}, A.
  2012, \aap, 539, A134

\bibitem[{{Bruzual} \& {Charlot}(2003)}]{bruzual03}
{Bruzual}, G., \& {Charlot}, S. 2003, \mnras, 344, 1000

\bibitem[{{Calzetti} {et~al.}(2007){Calzetti}, {Kennicutt}, {Engelbracht},
  {Leitherer}, {Draine}, {Kewley}, {Moustakas}, {Sosey}, {Dale}, {Gordon},
  {Helou}, {Hollenbach}, {Armus}, {Bendo}, {Bot}, {Buckalew}, {Jarrett}, {Li},
  {Meyer}, {Murphy}, {Prescott}, {Regan}, {Rieke}, {Roussel}, {Sheth}, {Smith},
  {Thornley}, \& {Walter}}]{calzetti07}
{Calzetti}, D., {Kennicutt}, R.~C., {Engelbracht}, C.~W., {et~al.} 2007, \apj,
  666, 870

\bibitem[{{Cardelli} {et~al.}(1989){Cardelli}, {Clayton}, \&
  {Mathis}}]{cardelli89}
{Cardelli}, J.~A., {Clayton}, G.~C., \& {Mathis}, J.~S. 1989, \apj, 345, 245

\bibitem[{{Colina}(1993)}]{colina93}
{Colina}, L. 1993, \apj, 411, 565

\bibitem[{{Colina} \& {Perez-Olea}(1992)}]{colina92}
{Colina}, L., \& {Perez-Olea}, D. 1992, \mnras, 259, 709

\bibitem[{{Colina} {et~al.}(1991){Colina}, {Sparks}, \& {Macchetto}}]{colina91}
{Colina}, L., {Sparks}, W.~B., \& {Macchetto}, F. 1991, \apj, 370, 102

\bibitem[{{Condon}(1992)}]{condon92}
{Condon}, J.~J. 1992, \araa, 30, 575

\bibitem[{{Condon} {et~al.}(1998){Condon}, {Cotton}, {Greisen}, {Yin},
  {Perley}, {Taylor}, \& {Broderick}}]{condon98}
{Condon}, J.~J., {Cotton}, W.~D., {Greisen}, E.~W., {et~al.} 1998, \aj, 115,
  1693

\bibitem[{{de Vaucouleurs} {et~al.}(1991){de Vaucouleurs}, {de Vaucouleurs},
  {Corwin}, {Buta}, {Paturel}, \& {Fouqu{\'e}}}]{devaucouleurs91}
{de Vaucouleurs}, G., {de Vaucouleurs}, A., {Corwin}, Jr., H.~G., {et~al.}
  1991, {Third Reference Catalogue of Bright Galaxies. Volume I: Explanations
  and references. Volume II: Data for galaxies between 0$^{h}$ and 12$^{h}$.
  Volume III: Data for galaxies between 12$^{h}$ and 24$^{h}$.}

\bibitem[{{D{\'{\i}}az-Santos} {et~al.}(2010){D{\'{\i}}az-Santos},
  {Alonso-Herrero}, {Colina}, {Packham}, {Levenson}, {Pereira-Santaella},
  {Roche}, \& {Telesco}}]{diaz-santos10}
{D{\'{\i}}az-Santos}, T., {Alonso-Herrero}, A., {Colina}, L., {et~al.} 2010,
  \apj, 711, 328

\bibitem[{{D{\'{\i}}az-Santos} {et~al.}(2008){D{\'{\i}}az-Santos},
  {Alonso-Herrero}, {Colina}, {Packham}, {Radomski}, \&
  {Telesco}}]{diaz-santos08}
---. 2008, \apj, 685, 211

\bibitem[{{Dickey} \& {Lockman}(1990)}]{dickley90}
{Dickey}, J.~M., \& {Lockman}, F.~J. 1990, \araa, 28, 215

\bibitem[{{Efstathiou} \& {Rowan-Robinson}(1995)}]{efstathiou95}
{Efstathiou}, A., \& {Rowan-Robinson}, M. 1995, \mnras, 273, 649

\bibitem[{{Efstathiou} {et~al.}(2000){Efstathiou}, {Rowan-Robinson}, \&
  {Siebenmorgen}}]{efstathiou00}
{Efstathiou}, A., {Rowan-Robinson}, M., \& {Siebenmorgen}, R. 2000, \mnras,
  313, 734

\bibitem[{{Efstathiou} \& {Siebenmorgen}(2009)}]{efstathiou09}
{Efstathiou}, A., \& {Siebenmorgen}, R. 2009, \aap, 502, 541

\bibitem[{{Genzel} {et~al.}(1998){Genzel}, {Lutz}, {Sturm}, {Egami}, {Kunze},
  {Moorwood}, {Rigopoulou}, {Spoon}, {Sternberg}, {Tacconi-Garman}, {Tacconi},
  \& {Thatte}}]{genzel98}
{Genzel}, R., {Lutz}, D., {Sturm}, E., {et~al.} 1998, \apj, 498, 579

\bibitem[{{Greenhouse} {et~al.}(1991){Greenhouse}, {Woodward}, {Thronson},
  {Rudy}, {Rossano}, {Erwin}, \& {Puetter}}]{greenhouse91}
{Greenhouse}, M.~A., {Woodward}, C.~E., {Thronson}, Jr., H.~A., {et~al.} 1991,
  \apj, 383, 164

\bibitem[{{Heesen} {et~al.}(2011){Heesen}, {Beck}, {Krause}, \&
  {Dettmar}}]{heesen11}
{Heesen}, V., {Beck}, R., {Krause}, M., \& {Dettmar}, R.-J. 2011, \aap, 535,
  A79

\bibitem[{{Helou} {et~al.}(1985){Helou}, {Soifer}, \&
  {Rowan-Robinson}}]{helou85}
{Helou}, G., {Soifer}, B.~T., \& {Rowan-Robinson}, M. 1985, \apjl, 298, L7

\bibitem[{{Hern{\'a}ndez-Garc{\'{\i}}a}
  {et~al.}(2013){Hern{\'a}ndez-Garc{\'{\i}}a}, {Gonz{\'a}lez-Mart{\'{\i}}n},
  {M{\'a}rquez}, \& {Masegosa}}]{hernandez-garcia13}
{Hern{\'a}ndez-Garc{\'{\i}}a}, L., {Gonz{\'a}lez-Mart{\'{\i}}n}, O.,
  {M{\'a}rquez}, I., \& {Masegosa}, J. 2013, \aap, 556, A47

\bibitem[{{Imanishi} \& {Nakanishi}(2013)}]{imanishi13}
{Imanishi}, M., \& {Nakanishi}, K. 2013, \aj, 146, 47

\bibitem[{{Ishihara} {et~al.}(2010){Ishihara}, {Onaka}, {Kataza}, {Salama},
  {Alfageme}, {Cassatella}, {Cox}, {Garc{\'{\i}}a-Lario}, {Stephenson},
  {Cohen}, {Fujishiro}, {Fujiwara}, {Hasegawa}, {Ita}, {Kim}, {Matsuhara},
  {Murakami}, {M{\"u}ller}, {Nakagawa}, {Ohyama}, {Oyabu}, {Pyo}, {Sakon},
  {Shibai}, {Takita}, {Tanab{\'e}}, {Uemizu}, {Ueno}, {Usui}, {Wada},
  {Watarai}, {Yamamura}, \& {Yamauchi}}]{ishihara10}
{Ishihara}, D., {Onaka}, T., {Kataza}, H., {et~al.} 2010, \aap, 514, A1

\bibitem[{{Iwasawa} {et~al.}(2011){Iwasawa}, {Sanders}, {Teng}, {U}, {Armus},
  {Evans}, {Howell}, {Komossa}, {Mazzarella}, {Petric}, {Surace}, {Vavilkin},
  {Veilleux}, \& {Trentham}}]{iwasawa11}
{Iwasawa}, K., {Sanders}, D.~B., {Teng}, S.~H., {et~al.} 2011, \aap, 529, A106

\bibitem[{{Kalberla} {et~al.}(2005){Kalberla}, {Burton}, {Hartmann}, {Arnal},
  {Bajaja}, {Morras}, \& {P{\"o}ppel}}]{kalberla05}
{Kalberla}, P.~M.~W., {Burton}, W.~B., {Hartmann}, D., {et~al.} 2005, \aap,
  440, 775

\bibitem[{{K{\"o}nig} {et~al.}(2013){K{\"o}nig}, {Aalto}, {Muller}, {Beswick},
  \& {Gallagher}}]{konig13}
{K{\"o}nig}, S., {Aalto}, S., {Muller}, S., {Beswick}, R.~J., \& {Gallagher},
  J.~S. 2013, \aap, 553, A72

\bibitem[{{Kotilainen} {et~al.}(2001){Kotilainen}, {Reunanen}, {Laine}, \&
  {Ryder}}]{kotilainen01}
{Kotilainen}, J.~K., {Reunanen}, J., {Laine}, S., \& {Ryder}, S.~D. 2001, \aap,
  366, 439

\bibitem[{{Leitherer} {et~al.}(1999){Leitherer}, {Schaerer}, {Goldader},
  {Gonz{\'a}lez Delgado}, {Robert}, {Kune}, {de Mello}, {Devost}, \&
  {Heckman}}]{leitherer99}
{Leitherer}, C., {Schaerer}, D., {Goldader}, J.~D., {et~al.} 1999, \apjs, 123,
  3

\bibitem[{{Mattila} \& {Meikle}(2001)}]{mattila01}
{Mattila}, S., \& {Meikle}, W.~P.~S. 2001, \mnras, 324, 325

\bibitem[{{Miralles-Caballero} {et~al.}(2011){Miralles-Caballero}, {Colina},
  {Arribas}, \& {Duc}}]{miralles-caballero11}
{Miralles-Caballero}, D., {Colina}, L., {Arribas}, S., \& {Duc}, P.-A. 2011,
  \aj, 142, 79

\bibitem[{{Moorwood} \& {Oliva}(1988)}]{moorwood88}
{Moorwood}, A.~F.~M., \& {Oliva}, E. 1988, \aap, 203, 278

\bibitem[{{Neff} {et~al.}(1990){Neff}, {Hutchings}, {Standord}, \&
  {Unger}}]{neff90}
{Neff}, S.~G., {Hutchings}, J.~B., {Standord}, S.~A., \& {Unger}, S.~W. 1990,
  \aj, 99, 1088

\bibitem[{{Olsson} {et~al.}(2010){Olsson}, {Aalto}, {Thomasson}, \&
  {Beswick}}]{olsson10}
{Olsson}, E., {Aalto}, S., {Thomasson}, M., \& {Beswick}, R. 2010, \aap, 513,
  A11

\bibitem[{{Osterbrock}(1989)}]{osterbrock89}
{Osterbrock}, D.~E. 1989, {Astrophysics of gaseous nebulae and active galactic
  nuclei}

\bibitem[{{Parra} {et~al.}(2007){Parra}, {Conway}, {Diamond}, {Thrall},
  {Lonsdale}, {Lonsdale}, \& {Smith}}]{parra07}
{Parra}, R., {Conway}, J.~E., {Diamond}, P.~J., {et~al.} 2007, \apj, 659, 314

\bibitem[{{Pereira-Santaella} {et~al.}(2011){Pereira-Santaella},
  {Alonso-Herrero}, {Santos-Lleo}, {Colina}, {Jim{\'e}nez-Bail{\'o}n},
  {Longinotti}, {Rieke}, {Ward}, \& {Esquej}}]{pereira-santaella11}
{Pereira-Santaella}, M., {Alonso-Herrero}, A., {Santos-Lleo}, M., {et~al.}
  2011, \aap, 535, A93

\bibitem[{{Perez-Olea} \& {Colina}(1996)}]{perez-olea96}
{Perez-Olea}, D.~E., \& {Colina}, L. 1996, \apj, 468, 191

\bibitem[{{P{\'e}rez-Torres} {et~al.}(2009{\natexlab{a}}){P{\'e}rez-Torres},
  {Alberdi}, {Colina}, {Torrelles}, {Panagia}, {Wilson}, {Kankare}, \&
  {Mattila}}]{perez-torres09a}
{P{\'e}rez-Torres}, M.~A., {Alberdi}, A., {Colina}, L., {et~al.}
  2009{\natexlab{a}}, \mnras, 399, 1641

\bibitem[{{P{\'e}rez-Torres} {et~al.}(2002){P{\'e}rez-Torres}, {Alberdi},
  {Marcaide}, {Guirado}, {Lara}, {Mantovani}, {Ros}, \&
  {Weiler}}]{perez-torres02a}
{P{\'e}rez-Torres}, M.~A., {Alberdi}, A., {Marcaide}, J.~M., {et~al.} 2002,
  \mnras, 335, L23

\bibitem[{{P{\'e}rez-Torres} {et~al.}(2010){P{\'e}rez-Torres}, {Alberdi},
  {Romero-Ca{\~n}izales}, \& {Bondi}}]{perez-torres10}
{P{\'e}rez-Torres}, M.~A., {Alberdi}, A., {Romero-Ca{\~n}izales}, C., \&
  {Bondi}, M. 2010, \aap, 519, L5+

\bibitem[{{P{\'e}rez-Torres} {et~al.}(2009{\natexlab{b}}){P{\'e}rez-Torres},
  {Romero-Ca{\~n}izales}, {Alberdi}, \& {Polatidis}}]{perez-torres09b}
{P{\'e}rez-Torres}, M.~A., {Romero-Ca{\~n}izales}, C., {Alberdi}, A., \&
  {Polatidis}, A. 2009{\natexlab{b}}, \aap, 507, L17

\bibitem[{{P{\'e}rez-Torres} {et~al.}(2005){P{\'e}rez-Torres}, {Alberdi},
  {Marcaide}, {Guerrero}, {Lundqvist}, {Shapiro}, {Ros}, {Lara}, {Guirado},
  {Weiler}, \& {Stockdale}}]{perez-torres05}
{P{\'e}rez-Torres}, M.~A., {Alberdi}, A., {Marcaide}, J.~M., {et~al.} 2005,
  \mnras, 360, 1055

\bibitem[{{Puxley} \& {Brand}(1994)}]{puxley94}
{Puxley}, P.~J., \& {Brand}, P.~W.~J.~L. 1994, \mnras, 266, 431

\bibitem[{{Risaliti} {et~al.}(2000){Risaliti}, {Gilli}, {Maiolino}, \&
  {Salvati}}]{risaliti00}
{Risaliti}, G., {Gilli}, R., {Maiolino}, R., \& {Salvati}, M. 2000, \aap, 357,
  13

\bibitem[{{Rosenberg} {et~al.}(2012){Rosenberg}, {van der Werf}, \&
  {Israel}}]{rosenberg12}
{Rosenberg}, M.~J.~F., {van der Werf}, P.~P., \& {Israel}, F.~P. 2012, \aap,
  540, A116

\bibitem[{{Sanders} {et~al.}(2003){Sanders}, {Mazzarella}, {Kim}, {Surace}, \&
  {Soifer}}]{sanders03}
{Sanders}, D.~B., {Mazzarella}, J.~M., {Kim}, D.-C., {Surace}, J.~A., \&
  {Soifer}, B.~T. 2003, \aj, 126, 1607

\bibitem[{{Skrutskie} {et~al.}(2006){Skrutskie}, {Cutri}, {Stiening},
  {Weinberg}, {Schneider}, {Carpenter}, {Beichman}, {Capps}, {Chester},
  {Elias}, {Huchra}, {Liebert}, {Lonsdale}, {Monet}, {Price}, {Seitzer},
  {Jarrett}, {Kirkpatrick}, {Gizis}, {Howard}, {Evans}, {Fowler}, {Fullmer},
  {Hurt}, {Light}, {Kopan}, {Marsh}, {McCallon}, {Tam}, {Van Dyk}, \&
  {Wheelock}}]{skrutskie06}
{Skrutskie}, M.~F., {Cutri}, R.~M., {Stiening}, R., {et~al.} 2006, \aj, 131,
  1163

\bibitem[{{Smartt}(2009)}]{smartt09}
{Smartt}, S.~J. 2009, \araa, 47, 63

\bibitem[{{Soifer} {et~al.}(2001){Soifer}, {Neugebauer}, {Matthews}, {Egami},
  {Weinberger}, {Ressler}, {Scoville}, {Stolovy}, {Condon}, \&
  {Becklin}}]{soifer01}
{Soifer}, B.~T., {Neugebauer}, G., {Matthews}, K., {et~al.} 2001, \aj, 122,
  1213

\bibitem[{{Soria} \& {Wu}(2003)}]{soria03}
{Soria}, R., \& {Wu}, K. 2003, \aap, 410, 53

\bibitem[{{Terlouw} \& {Vogelaar}(2012)}]{kapteyn12}
{Terlouw}, J.~P., \& {Vogelaar}, M.~G.~R. 2012, {Kapteyn Package, version 2.2},
  {Kapteyn Astronomical Institute}, Groningen, available from
  \url{http://www.astro.rug.nl/software/kapteyn/}

\bibitem[{{U} {et~al.}(2012){U}, {Sanders}, {Mazzarella}, {Evans}, {Howell},
  {Surace}, {Armus}, {Iwasawa}, {Kim}, {Casey}, {Vavilkin}, {Dufault},
  {Larson}, {Barnes}, {Chan}, {Frayer}, {Haan}, {Inami}, {Ishida},
  {Kartaltepe}, {Melbourne}, \& {Petric}}]{u12}
{U}, V., {Sanders}, D.~B., {Mazzarella}, J.~M., {et~al.} 2012, \apjs, 203, 9

\bibitem[{{V{\"a}is{\"a}nen} {et~al.}(2012){V{\"a}is{\"a}nen}, {Rajpaul},
  {Zijlstra}, {Reunanen}, \& {Kotilainen}}]{vaisanen12}
{V{\"a}is{\"a}nen}, P., {Rajpaul}, V., {Zijlstra}, A.~A., {Reunanen}, J., \&
  {Kotilainen}, J. 2012, \mnras, 420, 2209

\bibitem[{{Vanzi} \& {Rieke}(1997)}]{vanzi97}
{Vanzi}, L., \& {Rieke}, G.~H. 1997, \apj, 479, 694

\bibitem[{{Wilson} {et~al.}(2008){Wilson}, {Petitpas}, {Iono}, {Baker}, {Peck},
  {Krips}, {Warren}, {Golding}, {Atkinson}, {Armus}, {Cox}, {Ho}, {Juvela},
  {Matsushita}, {Mihos}, {Pihlstrom}, \& {Yun}}]{wilson08}
{Wilson}, C.~D., {Petitpas}, G.~R., {Iono}, D., {et~al.} 2008, \apjs, 178, 189

\bibitem[{{Yuan} {et~al.}(2010){Yuan}, {Kewley}, \& {Sanders}}]{yuan10}
{Yuan}, T.-T., {Kewley}, L.~J., \& {Sanders}, D.~B. 2010, \apj, 709, 884

\bibitem[{{Zaragoza-Cardiel} {et~al.}(2013){Zaragoza-Cardiel}, {Font-Serra},
  {Beckman}, {Blasco-Herrera}, {Garc{\'{\i}}a-Lorenzo}, {Camps},
  {Gonzalez-Martin}, {Ramos Almeida}, {Loiseau}, \&
  {Guti{\'e}rrez}}]{zaragoza-cardiel13}
{Zaragoza-Cardiel}, J., {Font-Serra}, J., {Beckman}, J.~E., {et~al.} 2013,
  \mnras, 432, 998

\end{thebibliography}

\end{document}